\documentclass[]{style}

\usepackage{amsmath}
\usepackage{amssymb}
\usepackage{tabularx}

\newcommand{\state}{\mathbf{x}}
\newcommand{\sequence}{s}
\newcommand{\lagtime}{\tau}
\newcommand{\pbcbox}{\mathbf{L}}

\definecolor{tumblue}{HTML}{0065BD}
\newcommand{\tumlogo}{%
  \begin{tikzpicture}[x=0.26mm,y=-0.26mm,baseline=(current bounding box.north)]
    \fill[tumblue]
      (28,0) -- (28,31) -- (36,31) -- (36,0) -- (73,0) --
      (73,38) -- (66,38) -- (66,7) -- (58,7) -- (58,38) --
      (51,38) -- (51,7) -- (43,7) -- (43,38) -- (21,38) --
      (21,7) -- (14,7) -- (14,38) -- (7,38) -- (7,7) --
      (0,7) -- (0,0) -- cycle;
  \end{tikzpicture}%
}

\author[1]{Weilong Chen}
\author[1]{Nuno Costa}
\author[1,2]{Julija Zavadlav}

\affiliation[1]{MFM Lab, TUM, Germany}
\affiliation[2]{AMC, TUM, Germany}

\correspondence{Julija Zavadlav (\email{julija.zavadlav@tum.de})}

\title{SupraTITO: Transferable Generative Molecular Dynamics for Supramolecular Systems}
\abstract{Peptide sequence governs both the structures formed through supramolecular
assembly and the dynamics by which they emerge, but predicting either requires
resolving slow collective processes among many interacting molecules.
Molecular dynamics (MD) provides microscopic insight into these processes, yet
the long timescales of assembly and the vast peptide sequence space make
systematic exploration computationally demanding. We introduce SupraTITO, a
transferable generative molecular dynamics (GenMD) framework for supramolecular
systems, demonstrated through peptide self-assembly. SupraTITO learns
transferable implicit transfer operators (TITO) conditioned on peptide
sequence, molecular topology, and periodic geometry, allowing configurations
to be propagated over physical intervals much longer than an MD integration
step. On a comprehensive dipeptide benchmark, SupraTITO generalizes to held-out
sequences and reproduces sequence-dependent structures and dynamics while
maintaining molecular integrity over long rollouts. Compared with direct
ensemble prediction trained on the same trajectory data, SupraTITO more
accurately reproduces assembly structures while also resolving their temporal
evolution. The learned dynamics generalize across peptide concentrations,
including dilute conditions not represented during training. These results
extend transferable GenMD to collective dynamics in periodic supramolecular
systems and provide a foundation for modeling related processes beyond peptide
assembly.
}

\begin{document}
\maketitle

\section{Introduction}
\label{sec:introduction}
Predicting how amino-acid sequence determines molecular structure and
function is a fundamental problem in molecular biophysics~\citep{dill2012protein}. Deep learning methods, most notably AlphaFold~\citep{jumper2021highly, abramson2024accurate} and related
approaches~\citep{baek2021accurate}, have transformed the prediction of folded
three-dimensional structures from sequence. Molecular behavior, however,
extends beyond a single static structure. Biomolecules can adopt multiple
conformations and transition between them, and these dynamics can be important for
biological function~\citep{henzler2007dynamic}. Moreover, many biological and
materials processes emerge through interactions among multiple molecules
rather than through the behavior of an isolated molecule
~\citep{lehn2002toward,whitesides2002self}. Understanding such systems requires
predicting not only molecular structures, but also how intermolecular
interactions give rise to collective organization and dynamics.

\begin{figure*}[!t]
  \centering
  \includegraphics[width=\textwidth]{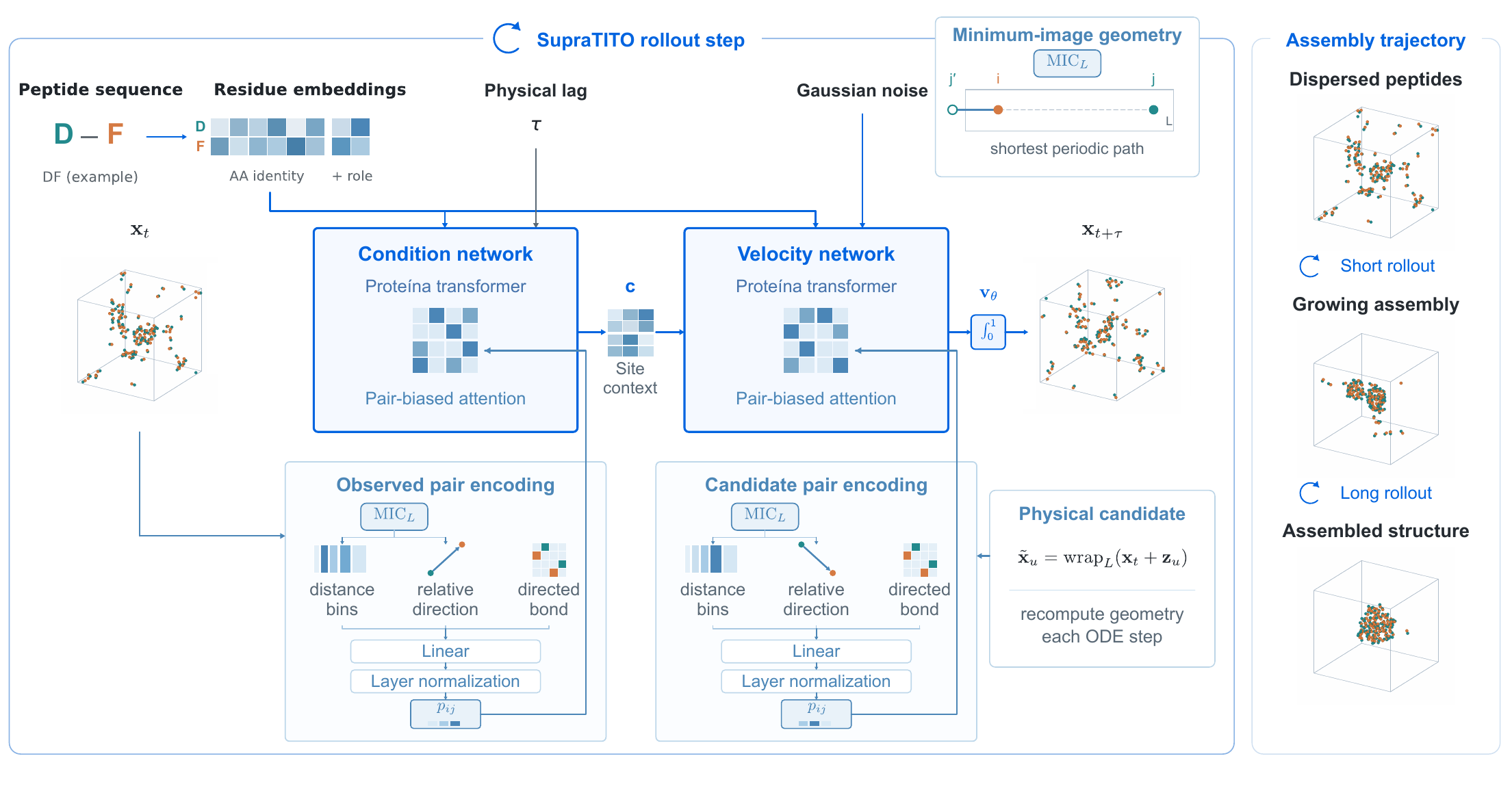}
  \caption{\textbf{SupraTITO architecture and generation of peptide assembly trajectories.}
  Peptide sequences are represented through learned amino-acid and terminal-role
  embeddings, illustrated for DF. The condition encoder combines these features
  with the current configuration \(\state_t\) and physical lag \(\lagtime\)
  to produce per-residue context for the velocity network. Both networks use
  pair representations that combine nonuniform minimum-image distance bins
  with directed intramolecular bond categories, providing learned attention
  biases for interactions within and between molecules. The condition encoder
  uses the observed geometry, whereas the velocity network recomputes pair
  distances from the intermediate configuration implied by
  \(\state_t+\mathbf{z}_u\) at each ODE evaluation. Integrating the velocity field from Gaussian noise
  produces a displacement; adding it to the current configuration and wrapping
  into the periodic box gives the next state. Repeated application with fresh
  noise generates an assembly trajectory. The DF snapshots illustrate
  assembly over multiple learned transitions. Green and coral distinguish residue
  positions, and sticks connect intramolecular pairs. Colored embedding blocks are schematic; molecular snapshots are centered for display.
  }
  \label{fig:patito-overview}
\end{figure*}

Peptide assembly is a prominent example of this broader supramolecular
prediction problem~\citep{yang2026past}. Peptides are chemically diverse,
modular building blocks whose sequences influence the formation of fibers,
nanotubes, sheets, and other supramolecular
morphologies~\citep{reches2003casting}. This versatility supports applications
in biomaterials, nanotechnology, and molecular engineering, but complicates
prediction: small sequence changes can alter intermolecular interactions,
while environmental conditions, including concentration, solvent, and
temperature, can redirect assembly pathways and
outcomes~\citep{kaygisiz2025context}. The combination of a rapidly growing sequence
space and many relevant physical conditions makes systematic experimental
exploration difficult.

Molecular dynamics (MD) provides a complementary route for investigating
assembly mechanisms and screening candidate sequences
computationally~\citep{frederix2015exploring,njirjak2024reshaping}. Assembly nevertheless
involves many interacting molecules and often develops through association,
reorganization, and other collective processes that are slow relative to the
microscopic MD integration step. Resolving these processes requires long
trajectories and, in many cases, multiple simulation replicates. Coarse-grained
force fields such as MARTINI~\citep{marrink2007martini} reduce this cost by representing groups of atoms
as effective interaction sites. Even with
coarse-graining, each combination of sequence and physicochemical conditions
must generally be simulated separately, making broad exploration expensive
~\citep{frederix2015exploring,shmilovich2020discovery}.

Generative molecular dynamics (GenMD) offers an alternative by learning
probabilistic surrogates of molecular
simulations~\citep{olsson2026generative}. Equilibrium generative models learn stationary distributions and generate samples from the corresponding ensembles
~\citep{noe2019boltzmann,zheng2024predicting,lewis2025scalable,chen2026coarse}, whereas dynamical GenMD methods learn time-dependent transitions between molecular
configurations
~\citep{schreiner2023implicit,klein2023timewarp,vlachas2022accelerated,
costa2024equijump}. In particular, implicit transfer
operator (ITO) learning models the finite-lag transition density
$p_{\lagtime}(\state_{t+\lagtime}\mid\state_t)$ using a conditional generative
model~\citep{schreiner2023implicit}. Repeated sampling from this density
generates stochastic trajectories without resolving every
intermediate MD step. When trained across related molecular systems, such models can also
generalize to systems not seen during training
~\citep{diez2026transferable,antoniadis2026platito}, allowing the cost of reference simulations and model training to be amortized across multiple targets.

Existing dynamical GenMD methods have primarily addressed the conformational
dynamics of individual molecules
~\citep{costa2024equijump,antoniadis2026platito}. Collective molecular
processes introduce additional challenges: many molecules move and reorganize,
intermolecular contacts continually form and break, and trajectories must
remain consistent with both molecular topology and the periodic simulation
domain. A transferable method must capture how molecular composition changes
these interactions while remaining stable during long recursive rollouts.

We introduce SupraTITO, a GenMD framework for \textbf{Supra}molecular dynamics
with \textbf{T}ransferable \textbf{I}mplicit \textbf{T}ransfer
\textbf{O}perators. SupraTITO learns finite-lag transition
densities conditioned on peptide sequence, molecular topology, and periodic
geometry using conditional flow matching~\citep{lipman2022flow}. A shared
molecular representation allows the learned dynamics to be applied to peptide
sequences excluded from training. On a comprehensive dipeptide benchmark,
the generated trajectories reproduce sequence-dependent assembly structures
and dynamics while preserving molecular integrity over long rollouts.
Compared with direct ensemble prediction trained on the same trajectory data,
SupraTITO more accurately reproduces the late-time structural ensemble. The
learned dynamics also generalize across peptide concentrations, including
lower concentrations where larger simulation volumes and longer molecular
encounter times can make direct MD particularly costly. Although demonstrated
here for peptide assembly, the formulation applies more broadly to collective
processes involving interacting molecules in periodic domains.

% Together, these results establish SupraTITO as a transferable generative framework for peptide assembly and extend GenMD to collective nonequilibrium dynamics in periodic many-molecule systems.

Our main contributions are:
\begin{itemize}[leftmargin=*,itemsep=2pt]

    \item \textbf{A GenMD framework for supramolecular dynamics.}
    SupraTITO extends GenMD beyond single-molecule conformational
    dynamics to collective processes. The formulation can, in principle, support related problems such as coassembly and other supramolecular processes, including micelles, self-assembled monolayers, phospholipid bilayers, viral capsids, and DNA origami.

    \item \textbf{Implicit transfer operators under periodic boundary conditions.}
    We extend ITO learning to periodic molecular systems using
    molecule-consistent minimum-image displacement targets, wrapped trajectory
    generation, and relative periodic geometry.

    \item \textbf{Evaluation of transfer across sequences and environments.}
    We systematically assess generalization across molecular sequences and physical conditions
    using structural and dynamical observables, intramolecular fidelity,
    direct ensemble prediction and training-sequence coverage experiments.

\end{itemize}

\section{Background and Preliminaries}
\label{sec:background}

\subsection{Molecular Dynamics and Transfer Operators}
\label{sec:background-md}

Molecular dynamics (MD) simulates the evolution of molecular systems by numerically integrating equations of motion under a potential energy model and specified simulation conditions~\citep{frenkel1997understanding}. Let $\mathbf{x}_t=(\mathbf{x}_{t,1},\ldots,\mathbf{x}_{t,N})\in\Omega\subseteq\mathbb{R}^{3N}$ denote the configuration of $N$ retained particles at physical time $t$. A trajectory $\{\mathbf{x}_t\}_{t\geq0}$ records the resulting configurational dynamics.

For time-homogeneous dynamics, the transition density
$p_\tau(\mathbf{x}'\mid\mathbf{x})$ describes the distribution of
configurations reached after a physical lag $\tau$, conditioned on
the current configuration $\mathbf{x}$. This density induces the transfer operator
\(\mathcal{P}_\tau\), which propagates a configuration density forward in time:
\begin{equation}
    \rho_{t+\tau}(\mathbf{x}')
    =(\mathcal{P}_\tau\rho_t)(\mathbf{x}')
    =\int_\Omega p_\tau(\mathbf{x}'\mid\mathbf{x})\,
        \rho_t(\mathbf{x})\,d\mathbf{x}.
    \label{eq:transfer-operator}
\end{equation}
Here, $\rho_t$ and $\rho_{t+\tau}$ denote the distributions of configurations
at times $t$ and $t+\tau$, respectively. The transition density describes evolution from a given configuration, whereas the transfer operator propagates an ensemble~\citep{prinz2011markov,nuske2014variational}.

Physical properties are expressed as expectations of observables over configuration space. For an observable $f:\Omega\rightarrow\mathbb{R}$,
\begin{equation}
    \langle f\rangle_t
    =\mathbb{E}_{\mathbf{x}\sim\rho_t}[f(\mathbf{x})]
    =\int_\Omega f(\mathbf{x})\,\rho_t(\mathbf{x})\,d\mathbf{x}.
    \label{eq:observable-expectation}
\end{equation}
Dynamical properties additionally depend on temporal correlations. For observables $f$ and $g$, the two-time correlation is
\begin{align}
    C_{f,g}(t,\tau)
    &=\mathbb{E}\!\left[f(\mathbf{x}_t)g(\mathbf{x}_{t+\tau})\right]
    \nonumber\\
    &=\int_\Omega\!\int_\Omega f(\mathbf{x})g(\mathbf{x}')\,
        p_\tau(\mathbf{x}'\mid\mathbf{x})\,\rho_t(\mathbf{x})\,
        d\mathbf{x}'\,d\mathbf{x}.
    \label{eq:lagged-correlation}
\end{align}
The transition density therefore determines both the evolution of
configurational statistics and the temporal correlations of molecular
observables.

\subsection{Conditional Flow Matching}
\label{sec:background-flow-matching}
Let \(p_0\) be a tractable reference
distribution on \(\mathbb{R}^d\) and \(p_1\) a target distribution. A
continuous normalizing flow transports samples from \(p_0\) to \(p_1\) through
the ordinary differential equation~\citep{chen2018neural}
\begin{equation}
    \frac{d\mathbf{z}_u}{du}
    =
    \mathbf{v}_\theta(\mathbf{z}_u,u),
    \qquad
    \mathbf{z}_0\sim p_0,
    \qquad
    u\in[0,1],
    \label{eq:flow-ode}
\end{equation}
where \(\mathbf{v}_\theta\) is a time-dependent vector field. The variable
\(u\) denotes generative flow time and is distinct from physical time \(t\)
and time lag \(\tau\).

Flow matching provides a simulation-free objective for training continuous
normalizing flows~\citep{lipman2022flow,liu2022flow,albergo2023stochastic}. Conditional flow matching constructs a tractable regression objective from
probability paths conditioned on target samples~\citep{tong2023improving,lipman2022flow}. Given a reference sample \(\mathbf{z}_0\), a target sample \(\mathbf{z}_1\sim p_1\), and
\(u\sim\mathcal{U}(0,1)\), the linear path
\begin{equation}
    \mathbf{z}_u
    =
    (1-u)\mathbf{z}_0+u\mathbf{z}_1
    \label{eq:linear-flow-path}
\end{equation}
has conditional velocity \(\mathbf{z}_1-\mathbf{z}_0\). The corresponding
objective is
\begin{equation}
    \mathcal{L}_{\mathrm{CFM}}(\theta)
    =
    \mathbb{E}_{\mathbf{z}_0,\mathbf{z}_1,u}
    \left[
        \left\|
            \mathbf{v}_\theta(\mathbf{z}_u,u)
            -(\mathbf{z}_1-\mathbf{z}_0)
        \right\|_2^2
    \right].
    \label{eq:conditional-flow-matching}
\end{equation}
The optimal regressor recovers the marginal velocity field associated with
the selected probability path. At inference, samples are generated by drawing
\(\mathbf{z}_0\sim p_0\) and integrating \cref{eq:flow-ode} from \(u=0\) to
\(u=1\).

\subsection{Implicit Transfer Operator Learning}
\label{sec:background-ito}

Implicit transfer operator (ITO) learning approximates finite-lag molecular transition densities using conditional generative models~\citep{schreiner2023implicit}:
\begin{equation}
    p_\theta(\mathbf{x}'\mid\mathbf{x},\tau)
    \approx
    p_\tau(\mathbf{x}'\mid\mathbf{x}).
    \label{eq:implicit-transition-model}
\end{equation}
The transfer operator is represented implicitly through a sampler of future configurations, rather than through an explicit transition matrix or basis expansion~\citep{prinz2011markov,nuske2014variational,husic2018markov}.

Training tuples $(\mathbf{x}_t,\mathbf{x}_{t+\tau},\tau)$ sampled from MD
trajectories instantiate $(\mathbf{x},\mathbf{x}')=(\mathbf{x}_t,
\mathbf{x}_{t+\tau})$ in \cref{eq:implicit-transition-model}. In a flow matching formulation, the current configuration $\mathbf{x}_t$ and lag $\tau$ provide the conditioning information, while $\mathbf{x}_{t+\tau}$ supplies a target sample. Sampling different lags during training allows one parameterization to represent a family of finite-time transition densities.

Starting from an initial configuration, a trajectory is generated by
repeatedly sampling from the learned density:
\begin{equation}
    \widehat{\mathbf{x}}_{(k+1)\tau}
    \sim p_\theta\!\left(\cdot\mid\widehat{\mathbf{x}}_{k\tau},\tau\right),
    \qquad k=0,1,\ldots.
    \label{eq:ito-recursive-rollout}
\end{equation}
Each application advances physical time by $\tau$, while the internal
generative flow spans $u\in[0,1]$. Conditioning each step only on the current
configuration defines a Markov process in the retained state. Its ability to
reproduce the reference dynamics depends on whether this state adequately
captures the relevant memory at the selected lag. Representations that retain
only positions or use coarse-graining can retain memory through unresolved
degrees of freedom~\citep{mori1965transport,klippenstein2021introducing}. Consequently,
consistency between a direct long-lag transition and a composition of shorter
transitions is not guaranteed and should be evaluated empirically~\citep{christians2026probabilistic}.

\section{Related Work}
\label{sec:related}

\paragraph{Generative molecular dynamics.}
Generative molecular dynamics learns molecular distributions to support
equilibrium sampling or the generation of dynamical
trajectories~\citep{olsson2026generative}. Molecular latent space simulators
learn slow collective variables, propagate dynamics in latent space, and
reconstruct continuous atomistic trajectories~\citep{sidky2020molecular}.
Autoregressive approaches propagate molecular states at a learned coarse
temporal resolution~\citep{vlachas2022accelerated,thiemann2026force}.
Other methods learn transitions between configurations. Timewarp learns
transferable normalizing flow proposals for molecular moves and combines them
with Markov chain Monte Carlo sampling~\citep{klein2023timewarp}. EquiJump uses
equivariant stochastic interpolants to generate transitions between protein
conformations~\citep{costa2024equijump}. MDGen instead learns a joint
distribution over trajectory segments for forward generation and other
conditional trajectory tasks~\citep{jing2024generative}. Within the ITO
framework, TITO extends finite-lag generation across molecular systems, and
PLaTITO incorporates protein language model embeddings to improve
generalization to unseen
proteins~\citep{diez2026transferable,antoniadis2026platito}. SupraTITO extends
this line of work from primarily single-molecule conformational dynamics to
peptide assembly in periodic many-molecule systems.

\paragraph{Machine learning for peptide assembly.}
Machine learning for peptide assembly has focused on predicting assembly propensity and navigating peptide sequence space. Supervised models relate sequence and molecular descriptors to
assembly propensity~\citep{batra2022machine}, while active learning
uses these predictions to guide further simulations and reduce
the number of candidates that require
evaluation~\citep{shmilovich2020discovery}. Generative design methods propose new
self-assembling sequences~\citep{njirjak2024reshaping}, with recent
work guiding generation toward desired aggregate morphologies
through isolated peptide descriptors~\citep{costa2026morphology}. These
approaches identify promising sequences and assembly outcomes. SupraTITO
instead generates the trajectories through which assemblies form and evolve,
providing complementary structural and dynamical information.

\section{SupraTITO: Generative Dynamics for Supramolecular Systems}
\label{sec:method}

Peptide assembly couples fixed molecular attributes, including sequence and
covalent topology, with evolving intermolecular organization. SupraTITO models
the distribution of molecular displacements over finite physical intervals
using a transformer architecture that integrates molecular features with periodic geometry. We describe the periodic formulation,
neural architecture, and training objective below.

\subsection{Implicit Transfer Operators under Periodic Boundary Conditions}
\label{sec:patito-operator}

Consider a periodic system containing \(M\) peptide molecules, with solvent
degrees of freedom excluded from the retained representation. Molecule \(m\)
has amino-acid sequence
\begin{equation}
    \sequence^{(m)}
    =
    \left(
        a^{(m)}_1,\ldots,a^{(m)}_{R_m}
    \right)
    \in
    \mathcal{A}^{R_m},
\end{equation}
where \(\mathcal{A}\) is the amino-acid alphabet, \(a_r^{(m)}\) is the identity
of residue \(r\), and \(R_m\) is the sequence length. The retained
representation contains \(N\) sites in total, with wrapped coordinates
\(\state_t\in\mathbb{R}^{N\times3}\), molecular graph \(\mathcal{G}\), and
orthorhombic box lengths \(\pbcbox=(L_x,L_y,L_z)\). Peptide self-assembly is the
homogeneous case in which all molecules share the same sequence.

% Periodic boundaries require additional care because applying the minimum-image convention independently to each site can assign different lattice-image branches to sites belonging to the same molecule. Although these independently wrapped displacements reconstruct the correct final configuration, their relative displacement can differ by an entire box vector, producing unphysical distortions of bonded geometry along the generative interpolation. We therefore construct a single consistent translational displacement for each molecule and represent internal motion separately.
 
SupraTITO represents each finite-lag transition as a displacement from the
observed configuration. Under periodic boundaries, applying the minimum-image
convention independently to bonded sites can select different periodic images,
introducing a box-length separation during flow interpolation. We therefore
decompose each molecular displacement into a common translation and an
internal change.

For the two-site per molecule representation used in this work, the sites of
molecule \(m\) are indexed by \(2m-1\) and \(2m\). The periodic bond vector and
geometric center are
\begin{equation}
    \mathbf b_t^{(m)}
    =
    \operatorname{MIC}_{\pbcbox}
    \left(
        \state_{t,2m}-\state_{t,2m-1}
    \right),
    \qquad
    \mathbf g_t^{(m)}
    =
    \state_{t,2m-1}+\frac{1}{2}\mathbf b_t^{(m)},
\end{equation}
where, component-wise,
\begin{equation}
    \left[
        \operatorname{MIC}_{\pbcbox}
        (\boldsymbol{\delta})
    \right]_k
    =
    \delta_k
    -
    L_k
    \operatorname{round}
    \left(
        \frac{\delta_k}{L_k}
    \right),
    \qquad k\in\{x,y,z\}.
\end{equation}
The molecular translation and internal change are
\begin{equation}
    \Delta\mathbf g_{t,\lagtime}^{(m)}
    =
    \operatorname{MIC}_{\pbcbox}
    \left(
        \mathbf g_{t+\lagtime}^{(m)}-\mathbf g_t^{(m)}
    \right),
    \qquad
    \Delta\mathbf b_{t,\lagtime}^{(m)}
    =
    \mathbf b_{t+\lagtime}^{(m)}-\mathbf b_t^{(m)}.
\end{equation}
These give the site displacements
\begin{equation}
    \mathbf d_{t,\lagtime,2m-1}
    =
    \Delta\mathbf g_{t,\lagtime}^{(m)}
    -\frac{1}{2}\Delta\mathbf b_{t,\lagtime}^{(m)},
    \qquad
    \mathbf d_{t,\lagtime,2m}
    =
    \Delta\mathbf g_{t,\lagtime}^{(m)}
    +\frac{1}{2}\Delta\mathbf b_{t,\lagtime}^{(m)}.
    \label{eq:patito-displacement}
\end{equation}
Collecting these site displacements gives
\(\mathbf d_{t,\lagtime}\in\mathbb{R}^{N\times3}\), the periodic displacement
from \(\state_t\) to \(\state_{t+\lagtime}\).
This construction assigns bonded sites a common molecular translation and
reconstructs the wrapped target configuration exactly.

Let
\(\mathcal{S}=\{\sequence^{(1)},\ldots,\sequence^{(M)}\}\) denote the
molecular sequences. SupraTITO approximates
\begin{equation}
    p_\theta(\mathbf d_{t,\lagtime}\mid
    \state_t,\mathcal S,\mathcal G,\lagtime,\pbcbox)
    \simeq
    p_{\lagtime}(\mathbf d_{t,\lagtime}\mid
    \state_t,\mathcal S,\mathcal G,\pbcbox).
    \label{eq:patito-density}
\end{equation}
A generated displacement is mapped back to periodic configuration space as
\begin{equation}
    \widehat{\state}_{t+\lagtime}
    =
    \operatorname{wrap}_{\pbcbox}
    \left(
        \state_t+\widehat{\mathbf{d}}_{t,\lagtime}
    \right).
    \label{eq:patito-reconstruction}
\end{equation}
A shared parameterization thus represents finite-lag transfer operators across
molecular sequences and lag times.

\subsection{Molecular Representation and Neural Architecture}
\label{sec:patito-architecture}

SupraTITO uses separate condition and velocity networks adapted from the Proteína
transformer architecture~\citep{geffner2025proteina,antoniadis2026platito}.
We extend this architecture to periodic many-molecule systems by introducing
directed molecular topology, minimum-image pair geometry, and relative
coordinate features. The condition network summarizes the observed
configuration and physical lag. The velocity network predicts a generative
vector field whose integration produces a displacement over that lag.

\paragraph{Node features.}
Each retained site \(i\) is described by its amino-acid type \(a_i\) and its
role \(r_i\) within the molecular graph:
\begin{equation}
    \mathbf f_i^{\mathrm{node}}
    =
    \mathbf{e}_{\mathrm{aa}}(a_i)\mathbin\Vert
    \mathbf{e}_{\mathrm{role}}(r_i).
    \label{eq:patito-node-features}
\end{equation}
The learned role embedding can distinguish positions such as terminal and
internal residues; for the dipeptides studied here, it distinguishes the
N-terminal and C-terminal sites. SupraTITO does not assign a single categorical
label to the complete sequence. A peptide is instead represented through the
types, roles, and connectivity of its constituent sites.

\paragraph{Pair features.}
For every ordered pair \((i,j)\), SupraTITO computes the minimum-image relative
vector, distance, and direction:
\begin{equation}
    \mathbf r_{ij}(\state)
    =
    \operatorname{MIC}_{\pbcbox}(\state_i-\state_j),
    \quad
    d_{ij}(\state)=\|\mathbf r_{ij}(\state)\|_2,
    \quad
    \operatorname{dir}_{ij}(\state)
    =
    \begin{cases}
        \mathbf r_{ij}(\state)/d_{ij}(\state), & i\ne j,\\
        \mathbf 0, & i=j.
    \end{cases}
    \label{eq:patito-relative-geometry}
\end{equation}
A directed bond category \(b_{ij}\) distinguishes nonbonded pairs and the two
directions along an intramolecular bond. The complete pair feature is
\begin{equation}
    \mathbf f_{ij}^{\mathrm{pair}}(\state)
    =
    \operatorname{onehot}(\operatorname{bin}(d_{ij}(\state)))
    \mathbin\Vert
    \operatorname{dir}_{ij}(\state)
    \mathbin\Vert
    \operatorname{onehot}(b_{ij}).
    \label{eq:patito-pair-features}
\end{equation}
Nonuniform distance categories provide greater resolution for bonded and
short-range intermolecular geometry while retaining coverage of more distant
pairs. A learned projection maps these features and applies layer normalization
(LN) to the pair representation
\begin{equation}
    \mathbf p_{ij}(\state)
    =
    \operatorname{LN}(W_{\mathrm{pair}}
    \mathbf f_{ij}^{\mathrm{pair}}(\state)).
\end{equation}
Here, \(\mathbf p(\state)=(\mathbf p_{ij}(\state))_{i,j=1}^{N}\) denotes the
complete pair representation.

\paragraph{Condition and velocity networks.}
The condition network combines a sinusoidal lag embedding with the molecular
features:
\begin{equation}
    \boldsymbol{\eta}^{\mathrm{cond}}_i
    =
    W_{\mathrm{cond}}
    [\mathbf{e}_{\lagtime}(\lagtime)\mathbin\Vert
    \mathbf f_i^{\mathrm{node}}].
\end{equation}
It jointly processes all site inputs and pair representations to produce one
context vector per site:
\begin{equation}
    \mathbf c_{t,\lagtime}
    =
    \left(\mathbf c_{t,\lagtime,i}\right)_{i=1}^{N}
    =
    f_{\mathrm{cond}}
    \left(
        \left(\boldsymbol{\eta}^{\mathrm{cond}}_i\right)_{i=1}^{N},
        \mathbf p(\state_t)
    \right).
    \label{eq:patito-condition}
\end{equation}
Absolute coordinates are not provided to the condition network. The observed
structure enters through translation-invariant periodic pair geometry.

The velocity network receives the intermediate displacement \(\mathbf z_u\),
the site contexts \(\mathbf c_{t,\lagtime}\),
the molecular features, and the flow-time embedding \(\mathbf{e}_u(u)\).
The physical lag therefore enters the velocity network through
\(\mathbf c_{t,\lagtime}\), whereas \(u\) is provided directly. At each flow
time, \(\mathbf z_u\) defines the intermediate configuration
\begin{equation}
    \state_u^{\mathrm{flow}}
    =
    \operatorname{wrap}_{\pbcbox}(\state_t+\mathbf z_u).
    \label{eq:patito-flow-configuration}
\end{equation}
The velocity network computes pair geometry from
\(\state_u^{\mathrm{flow}}\), allowing it to observe the bond lengths,
contacts, and periodic neighborhoods implied by the current displacement.

Within each transformer layer, pair features affect both the attention weights
and the information passed between sites:
\begin{align}
    A_{ij}^{(h)}
    &=
    \frac{\langle\mathbf q_i^{(h)},\mathbf k_j^{(h)}\rangle}{\sqrt{d_h}}
    +g_b^{(h)}(\mathbf p_{ij}),\\
    \alpha_{ij}^{(h)}
    &=\operatorname{softmax}_j(A_{ij}^{(h)}),\\
    \mathbf o_i^{(h)}
    &=
    \sum_j\alpha_{ij}^{(h)}
    [\mathbf v_j^{(h)}+g_v^{(h)}(\mathbf p_{ij})].
    \label{eq:patito-pair-attention}
\end{align}
The pair bias \(g_b^{(h)}\) modulates how strongly each site attends to other
sites, while \(g_v^{(h)}\) contributes pair-specific geometric and topological
information to the update. After
repeated transformer updates, the velocity network returns one
three-dimensional generative flow velocity per site. These outputs describe
motion along the generative path and are not physical MD velocities.

\subsection{Training and Trajectory Generation}
\label{sec:patito-training}

For each time-lagged MD pair, we instantiate the linear path in
\cref{eq:linear-flow-path} with
\(\mathbf z_1=\mathbf d_{t,\lagtime}\) and add a small Gaussian perturbation.
Specifically, we sample \(\mathbf z_0,\boldsymbol\epsilon\sim
\mathcal N(\mathbf 0,\mathbf I_{3N})\) and
\(u\sim\mathcal U(0,1)\), and construct
\begin{equation}
    \mathbf z_u
    =
    (1-u)\mathbf z_0
    +u\mathbf d_{t,\lagtime}
    +\sigma\boldsymbol\epsilon.
    \label{eq:patito-flow-path}
\end{equation}
The velocity network is trained using
\begin{equation}
    \mathcal L_{\mathrm{SupraTITO}}
    =
    \mathbb E\!\left[
        \frac{1}{3\sum_i w_i}
        \sum_{i=1}^{N}
        w_i
        \left\|
        \mathbf v_{\theta,i}(\mathbf z_u,u;\mathbf c_{t,\lagtime})
        -
        (\mathbf d_{t,\lagtime,i}-\mathbf z_{0,i})
        \right\|_2^2
    \right],
    \label{eq:patito-loss}
\end{equation}
where \(w_i\) masks padded sites. Training uses no assembly labels, force
targets, or auxiliary bond-length losses.

During inference, the condition network is evaluated once for the current
configuration. The resulting site contexts remain fixed while the flow ODE is
integrated from Gaussian noise to a sampled displacement. The resulting configuration
conditions the next finite-lag transition, and fresh Gaussian noise is drawn at
each step. Repeating this procedure generates a stochastic molecular
trajectory.

\paragraph{Symmetry.}
The architecture is permutation equivariant under a consistent reindexing of
sites and pair features. Minimum-image geometry is independent of the periodic
image used to represent each site, while relative geometry in the condition
network and displacement coordinates in the velocity network make the
architecture invariant to global translation. Common wrapped translations of
training pairs are retained as numerical augmentation. Because Cartesian
directions are used directly, the architecture is not exactly rotation
equivariant.

\section{Results}
\label{sec:results}

\paragraph{Dataset.}
We train and evaluate SupraTITO on a coarse-grained MD dataset containing all
$20^2=400$ ordered dipeptide sequences. Each \(1.25\,\mu\mathrm{s}\)
simulation contains 100 peptides at a concentration of approximately
\(100\,\mathrm{mM}\) and uses the Martini 2.1 force field
\citep{marrink2007martini}. We use neutral termini,
$\beta$-sheet backbone assignment, standard Martini water, and neutralizing
counterions, following the screening parameterization of
He et al.~\citep{he2026peptides}. Configurations are stored every
$2.5\,\mathrm{ps}$.

From each saved configuration, SupraTITO retains the MARTINI backbone (BB) bead of each amino-acid residue, giving two sites per
peptide and 200 sites per configuration. We use a reverse-pair-grouped split
with 320 training sequences and 80 held-out sequences. The split balances
amino-acid coverage and the distribution of late-time MD assembly behavior,
while assigning both orderings of each heterodipeptide to the same partition. Training lags are sampled uniformly from 1 to 1000 saved frames. No assembly
labels or derived observables are used during training. Further simulation,
split, and rollout details are provided in
\cref{app:simulation-details,app:splits,app:transfer-panel,app:rollout-settings}.

\paragraph{Metrics.}
We compare SupraTITO with MD using observables that describe complementary aspects
of peptide assembly. Solvent-accessible surface area (SASA) measures molecular
exposure to solvent. Peptide contact count and largest-cluster fraction quantify local
association and collective aggregation, while the radius of gyration (\(R_g\)), nematic
order, and relative shape anisotropy describe the size, orientational order,
and morphology of the largest aggregate. Bond-length distributions assess the
internal structure of individual peptides. We additionally compare the time evolution of SASA and
largest-cluster fraction, peptide-pair contact persistence, and
largest-cluster-fraction autocorrelation functions. Complete definitions and periodic-boundary treatments are provided
in \cref{app:metrics}. Inference uses a lag of 1000 saved frames. Generated
observables are averaged over five independent rollouts from the same initial
configuration and compared with the corresponding MD trajectory.
Computational costs are summarized in \cref{app:computational-cost}.

\subsection{SupraTITO transfers assembly dynamics across peptide sequences}
\label{sec:results-unseen}

\begin{figure*}[!htbp]
  \centering
  \includegraphics[width=0.96\textwidth]{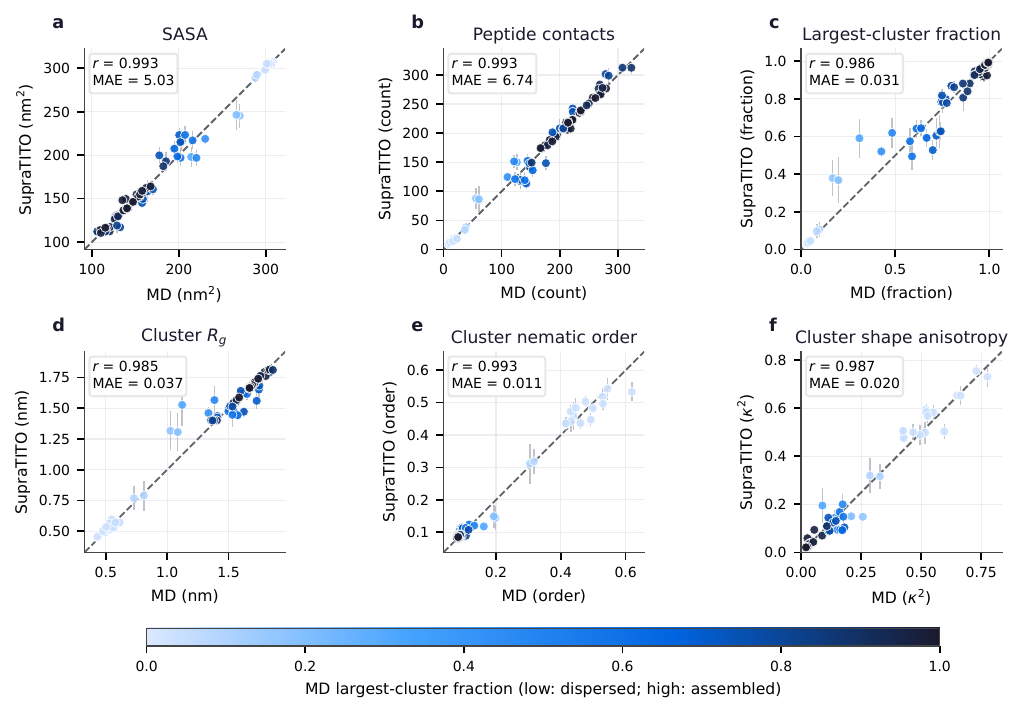}
  \caption{\textbf{Transfer to held-out dipeptide sequences.}   Markers show means across five independent SupraTITO rollouts, and vertical
error bars show one standard deviation of their late-window means. MD values
are averaged over the final \(125\,\mathrm{ns}\). Each point represents one
ordered sequence and is colored by its MD largest-cluster fraction, providing
a common measure of aggregation across peptide chemistries. Dashed lines
indicate exact agreement. Panel annotations report Pearson correlation
(\(r\)) and mean absolute error (MAE).}
  \label{fig:unseen-transfer}
\end{figure*}

\begin{figure*}[!htbp]
  \centering
  \includegraphics[width=0.92\textwidth]{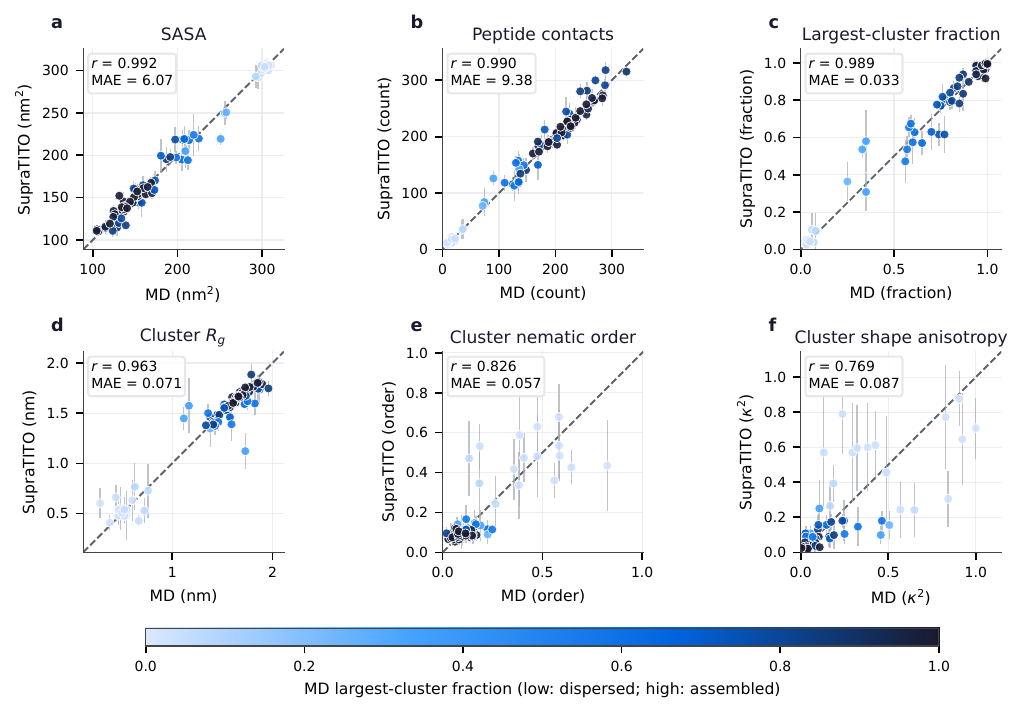}
  \caption{\textbf{Final-frame transfer to held-out dipeptide sequences.}
  SupraTITO is compared with MD using the instantaneous final configuration of
  each held-out trajectory. Each point is one ordered sequence,
  vertical bars show the rollout standard deviation, and
  color denotes the MD final-frame largest-cluster fraction. Dashed lines
  indicate exact agreement. Relative to the late-window averages in
  \cref{fig:unseen-transfer}, the final-frame values retain the principal
  sequence trends but exhibit greater dispersion because no temporal averaging
  suppresses structural fluctuations or differences between metastable
  realizations.}
  \label{fig:unseen-transfer-final}
\end{figure*}

SupraTITO reproduces the strong sequence dependence of peptide assembly.
Generated late-window averages closely match the corresponding MD values for
all six observables, across dispersed systems, compact aggregates, and
intermediate structures (\cref{fig:unseen-transfer}). The largest differences
and rollout variability occur mainly for intermediate systems, where
association and dissociation continue within the averaging window and finite
trajectories may occupy different partially assembled structures. Late-window
averaging reduces the effect of these fluctuations while retaining systematic
differences among sequences. Measurements from the final frame vary more
widely, particularly for orientational order and aggregate shape, but preserve
the same overall trends (\cref{fig:unseen-transfer-final}).

Generalization to held-out sequences suggests that interactions learned across
the training systems can be reused in new residue combinations. The
sequence-coverage experiment below examines how this transfer depends on the
diversity of the training set.

SupraTITO also preserves the internal structure of individual peptides
throughout the generated trajectories. Its bond-length distributions closely
match the MD references (\cref{app:heldout-bonds}), and indexed
intramolecular connections remain intact in the final configurations
(\cref{app:heldout-xy}). Thus, agreement in collective structure is accompanied
by molecular fidelity throughout long recursive rollouts.

\begin{figure*}[!htbp]
  \centering
  \includegraphics[width=0.96\textwidth]{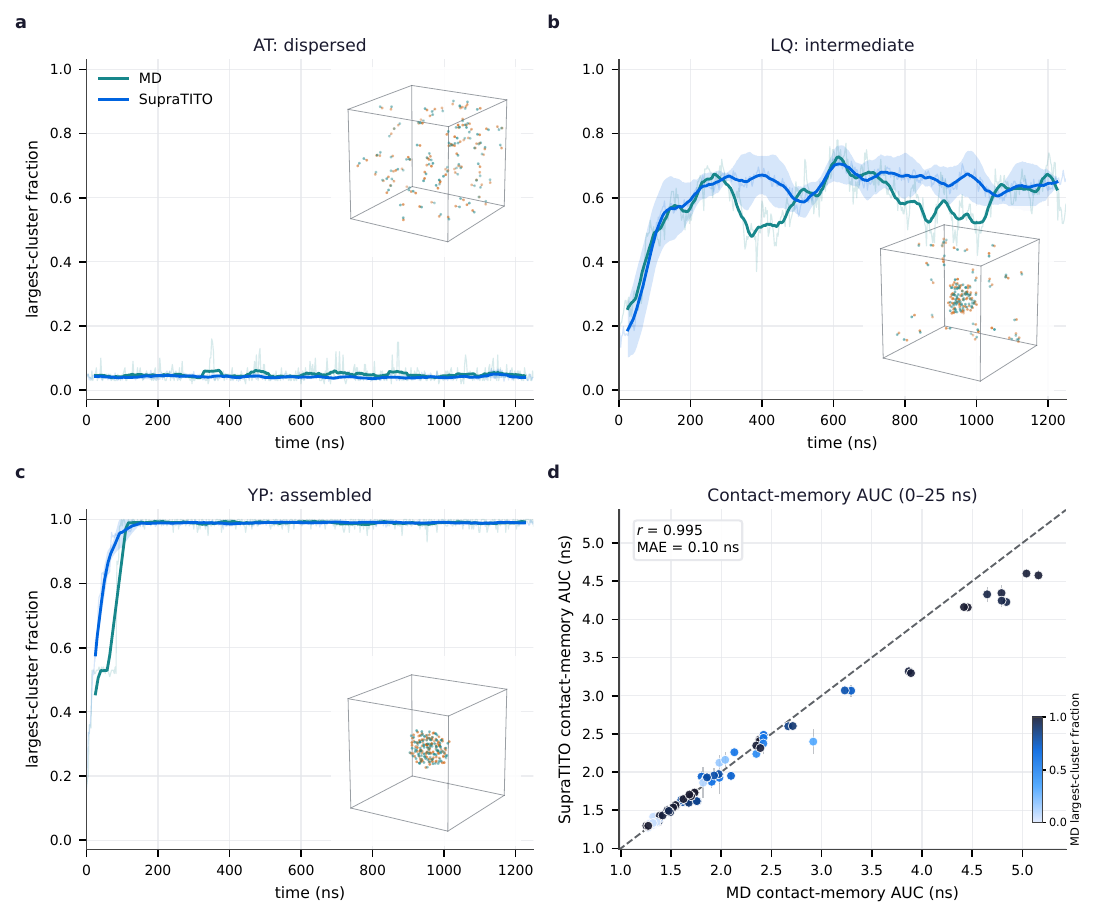}
  \caption{\textbf{Dynamical fidelity on held-out sequences.}
  \textbf{a--c}, largest-cluster fraction for representative dispersed,
  intermediate, and assembled sequences. Green curves show MD and blue curves
  show SupraTITO. Faint green curves show raw MD values, faint blue curves show
  framewise means across rollouts, solid curves show 21-frame moving averages,
  and blue bands show one standard deviation across smoothed SupraTITO rollouts. Insets
  show the final MD configurations, with teal and orange denoting the first
  and second residue sites of each peptide. \textbf{d}, baseline-corrected
  contact-persistence area under the curve over $0$--$25\,\mathrm{ns}$. Markers
  and vertical error bars show the mean and one standard deviation across
  SupraTITO rollouts and are colored by the MD late-window largest-cluster
  fraction.}
  \label{fig:dynamics-bonds}
\end{figure*}

\begin{figure*}[!htbp]
  \centering
  \includegraphics[width=0.92\textwidth,height=0.82\textheight,keepaspectratio]{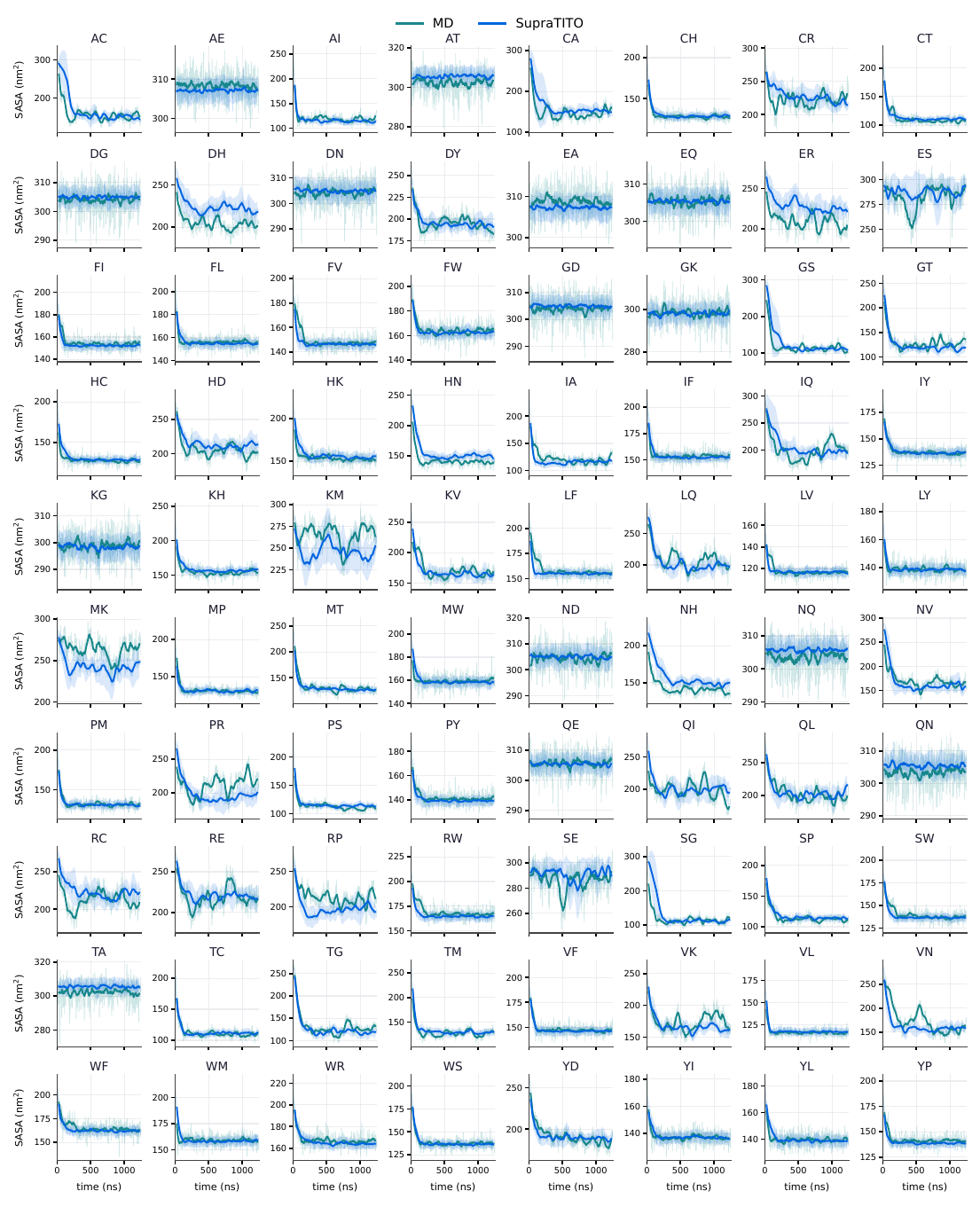}
  \caption{\textbf{SASA trajectories for all held-out sequences.} Each panel
  compares MD (green) and SupraTITO (blue) for one ordered sequence. Faint
  green curves show raw MD values, faint blue curves show framewise means
  across rollouts, solid curves show 21-frame moving averages, and blue bands
  show one standard deviation across rollouts. Each panel retains its own SASA
  scale.}
  \label{fig:all-unseen-sasa}
\end{figure*}

\begin{figure*}[!htbp]
  \centering
  \includegraphics[width=0.92\textwidth,height=0.82\textheight,keepaspectratio]{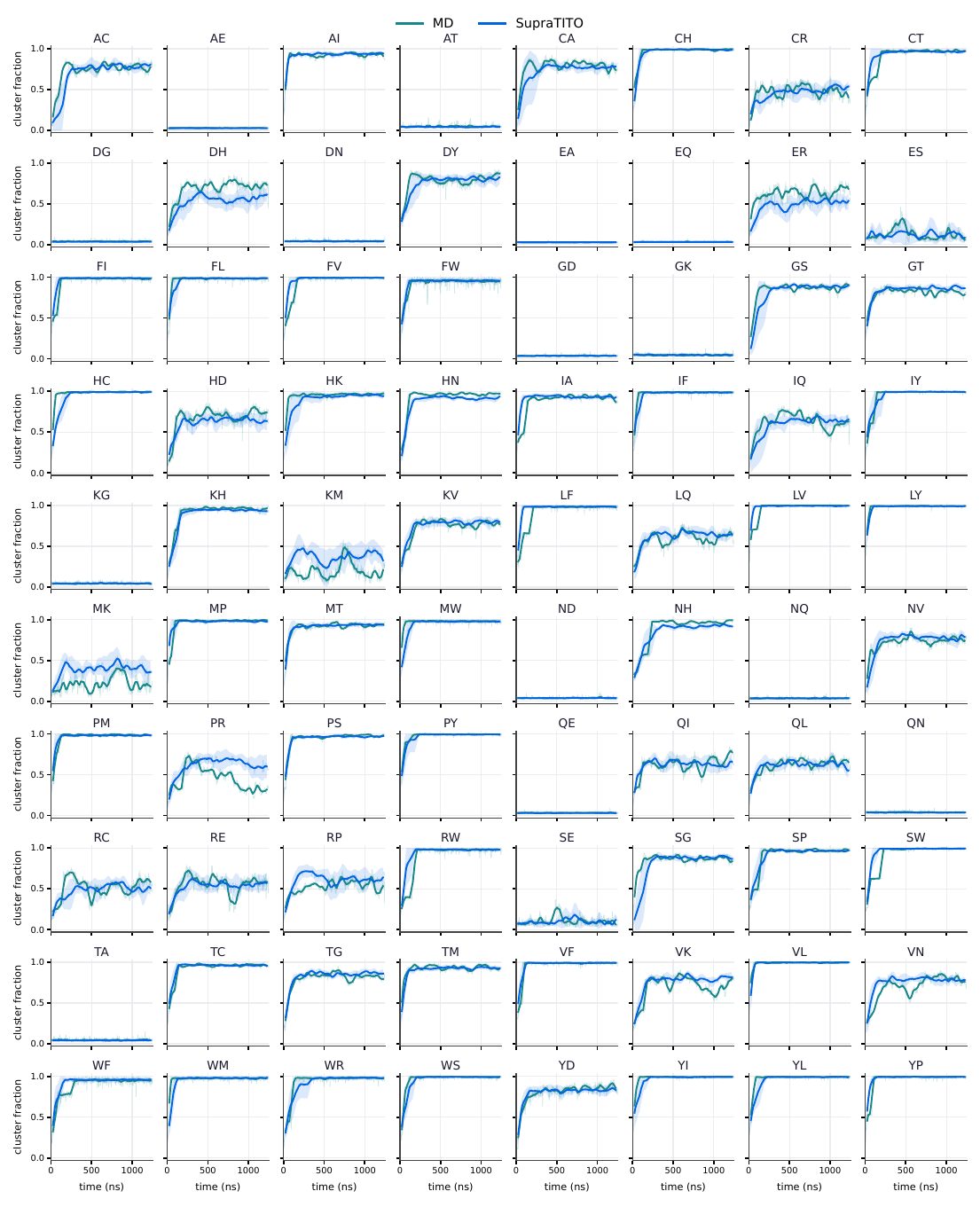}
  \caption{\textbf{Largest-cluster-fraction trajectories for all held-out
  sequences.} Green curves show MD and blue curves show SupraTITO. Faint curves
  show raw MD values and framewise SupraTITO means, solid curves show 21-frame
  moving averages, and blue bands show one standard deviation across rollouts.
  The observable is computed
  from the PBC-aware peptide-center contact graph.}
  \label{fig:all-unseen-cluster-fraction}
\end{figure*}

Beyond these structural comparisons, representative trajectories provide a
direct view of the assembly dynamics. Dispersed (AT), intermediate (LQ), and
assembled (YP) systems exhibit distinct cluster-growth behavior, and SupraTITO
reproduces both their overall evolution and temporal fluctuations
(\cref{fig:dynamics-bonds}a--c). Corresponding trajectories for every held-out
sequence are reported in
\cref{fig:all-unseen-sasa,fig:all-unseen-cluster-fraction}, with MD and SupraTITO
snapshots for the representative systems in
\cref{fig:trajectory-snapshots-at,fig:trajectory-snapshots-lq,fig:trajectory-snapshots-yp}.

\begin{figure*}[!htbp]
  \centering
  \includegraphics[width=0.92\textwidth,height=0.82\textheight,keepaspectratio]{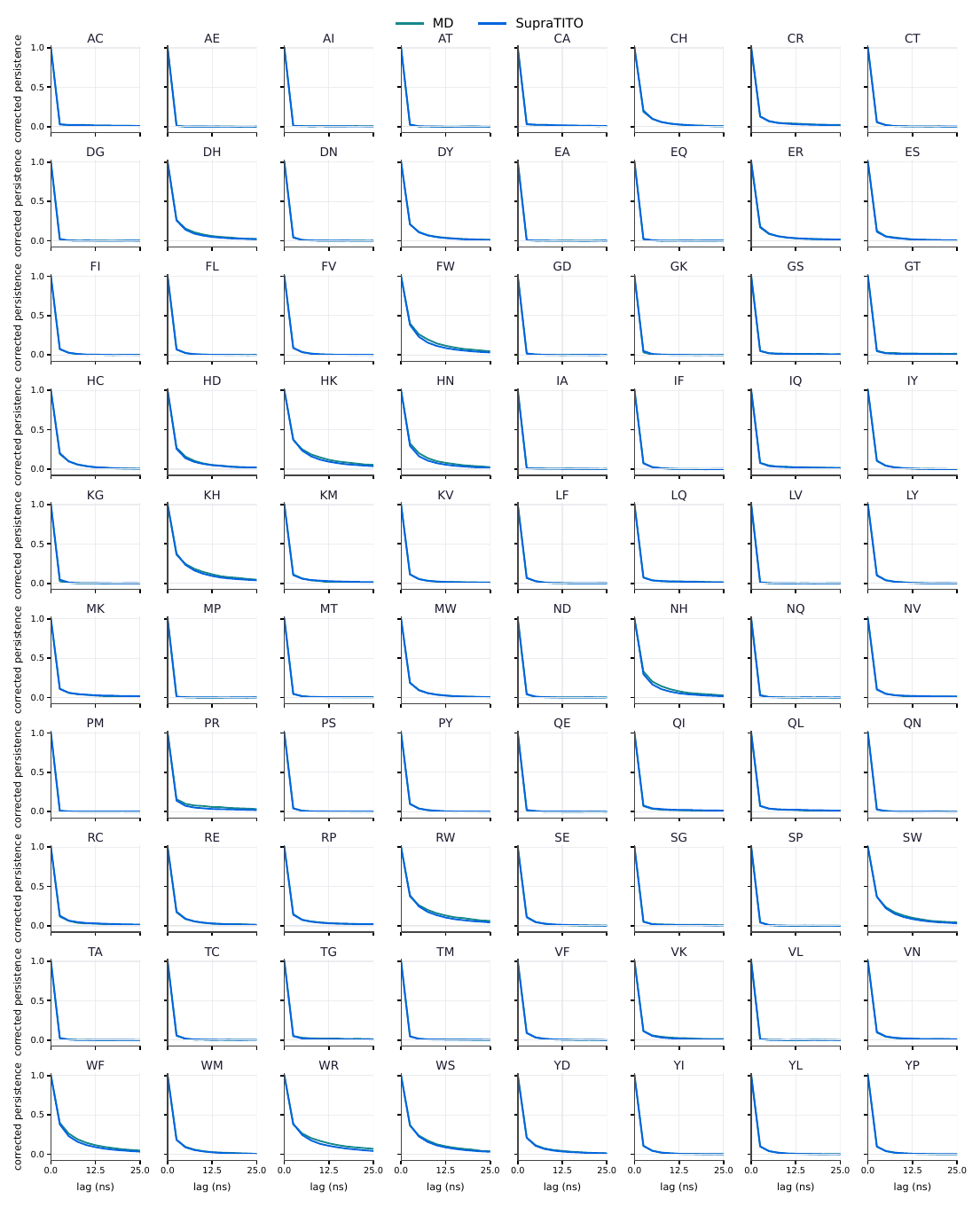}
  \caption{\textbf{Baseline-corrected contact persistence for all held-out
  sequences.} Each panel compares peptide-pair contact memory from MD and
  SupraTITO over the final $250\,\mathrm{ns}$, with MD in green and SupraTITO
  in blue. The sequence- and
  trajectory-specific contact-occupancy baseline is removed before averaging
  the generated curves. The blue band shows one standard deviation across
  rollouts. Persistence is evaluated for lags up to $25\,\mathrm{ns}$ using the
  peptide-center contact definition in \cref{eq:contact-graph}.}
  \label{fig:all-unseen-contact-persistence}
\end{figure*}

\begin{figure*}[!htbp]
  \centering
  \includegraphics[width=0.92\textwidth,height=0.82\textheight,keepaspectratio]{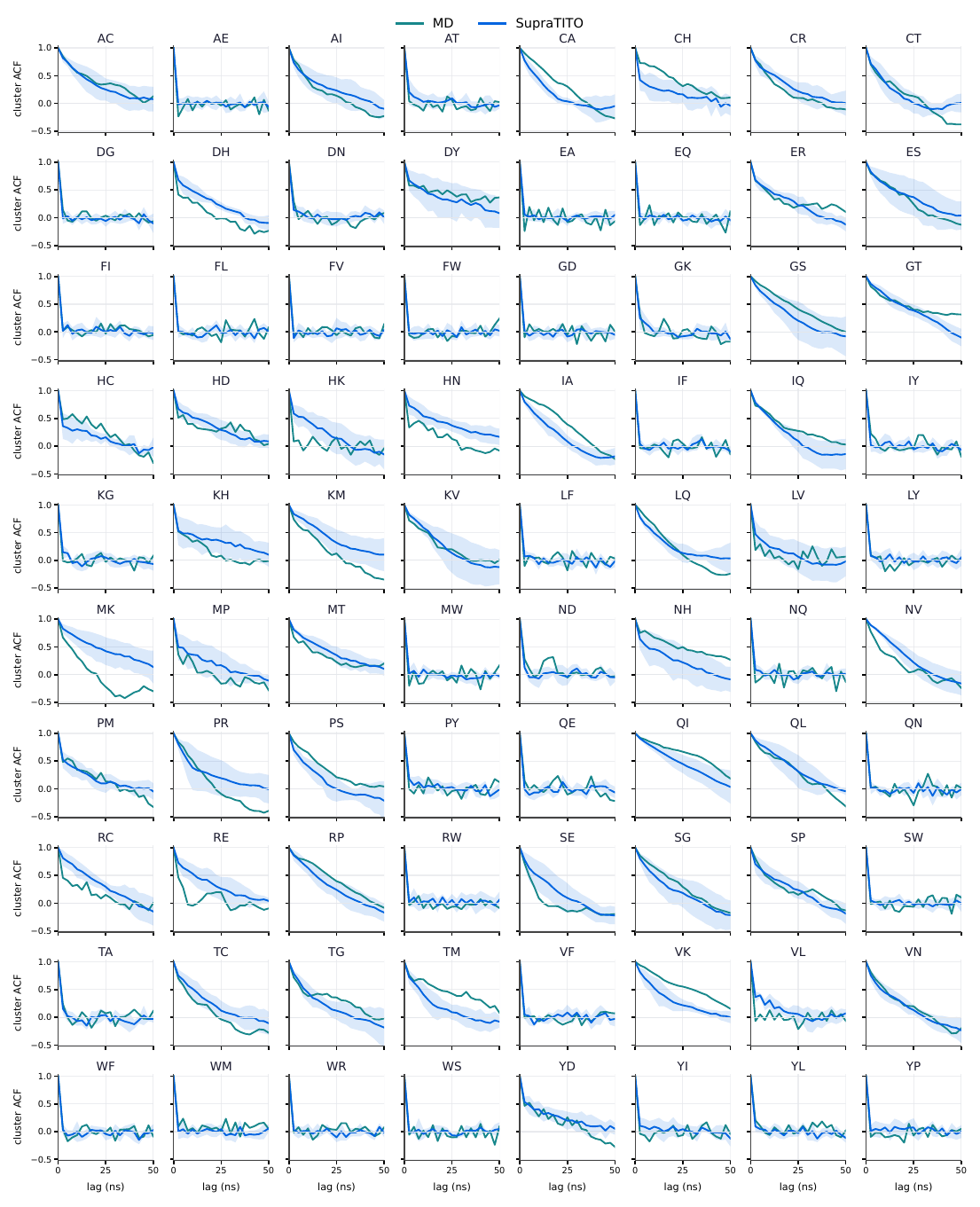}
  \caption{\textbf{Largest-cluster-fraction autocorrelation for all held-out
  sequences.} Each panel compares MD (green) and SupraTITO (blue)
  autocorrelations computed over the final $250\,\mathrm{ns}$, using offsets up
  to $50\,\mathrm{ns}$. The blue band shows one standard deviation across
  independently evaluated rollouts.
  These curves complement the contact-memory comparison in
  \cref{fig:dynamics-bonds}d.}
  \label{fig:all-unseen-acf}
\end{figure*}

The processed trajectories contain wrapped coordinates without molecular
winding information, which prevents an unambiguous estimate of long-time
diffusion. We therefore characterize relative assembly dynamics through
contact memory. For each sequence and trajectory, we remove the chance-contact
baseline set by its mean contact occupancy from the intermittent persistence
of peptide-pair contacts. Integrating the corrected curve over the first
$25\,\mathrm{ns}$ measures how strongly peptide-pair contacts persist above
this baseline. SupraTITO reproduces its sequence-dependent
variation, with a modest underestimation for some of the longest-lived contacts
(\cref{fig:dynamics-bonds}d). Complete corrected persistence curves and
windowed autocorrelation functions are provided in
\cref{fig:all-unseen-contact-persistence,fig:all-unseen-acf}.

\begin{figure*}[!htb]
  \centering
  \includegraphics[width=0.90\textwidth]{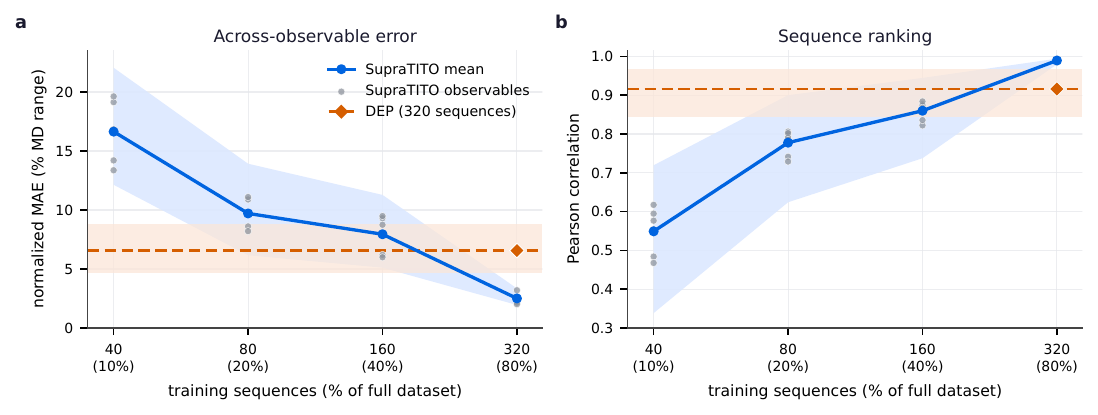}
  \caption{\textbf{Transfer improves with training-sequence coverage.}
  SupraTITO is trained on nested subsets of 40, 80, 160, and 320 ordered
  sequences and evaluated on a common held-out set. Percentages denote the
  fraction of the complete sequence corpus used for training.
  \textbf{a}, mean absolute error for each observable, normalized by the range
  of its MD late-window means. \textbf{b}, Pearson correlation between
  generated and MD late-window means. Gray markers show the individual
  SupraTITO observables and blue markers show their arithmetic mean. The
  orange dashed lines and diamonds show the full-data DEP benchmark described
  in \cref{sec:results-direct}; orange and blue bands are 95\% bootstrap
  confidence intervals. The horizontal DEP references do not represent a
  coverage sweep.}
  \label{fig:sequence-coverage}
\end{figure*}

\paragraph{Dependence on training-sequence coverage.}
We further train SupraTITO on nested subsets containing 40, 80, and 160 ordered
sequences and compare them with the model trained on the complete training
partition. All models are evaluated on the same held-out sequences. Errors
decrease and correlations with MD increase across all six observables as
training-sequence coverage grows (\cref{fig:sequence-coverage}). This
improvement continues after every amino-acid type is represented in the
training set. Transfer therefore depends not only on amino-acid coverage, but
also on exposure to diverse residue combinations and intermolecular
environments. The nested subset construction is described in
\cref{app:coverage-splits}.

\subsection{Finite-lag learning outperforms direct ensemble prediction}
\label{sec:results-direct}

We compare SupraTITO with a sequence-conditioned direct ensemble prediction (DEP)
baseline trained on the same training sequences and MD trajectories. DEP uses
the same residue features, directed molecular topology, and periodic pair
geometry as SupraTITO, but directly learns the late-structure distribution
\begin{equation}
    p_\phi\!\left(
        \state_{\mathrm{late}}
        \mid
        \sequence,\mathcal{G},\pbcbox
    \right).
\end{equation}
It therefore generates independent assembly configurations rather than a
trajectory ordered in time. Training targets are sampled from the final
$125\,\mathrm{ns}$ of each MD trajectory, where the dominant assembly
structures are comparatively stable. Sampling this late window provides a
structurally varied target ensemble; using only one endpoint from each
trajectory would provide too few configurations for conditional flow matching.
Implementation and evaluation details are provided in
\cref{app:direct-supplement}.

\begin{figure*}[!htb]
  \centering
  \begin{minipage}[t]{0.485\textwidth}
    \vspace{0pt}
    \hspace{0.12\linewidth}{\sffamily\bfseries\fontsize{8.8}{10}\selectfont a}\par
    \vspace{0.15em}
    \centering
    \includegraphics[width=\linewidth]{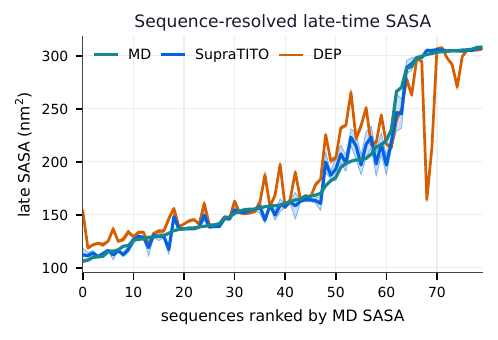}
  \end{minipage}\hfill
  \begin{minipage}[t]{0.485\textwidth}
    \vspace{0pt}
    {\sffamily\bfseries\fontsize{8.8}{10}\selectfont b}\par

    \vspace{0.15em}
    \centering
    \scriptsize
    \setlength{\tabcolsep}{1.7pt}
    \renewcommand{\arraystretch}{1.24}
    \begin{tabular}{@{}lcccc@{}}
      \toprule
      & \multicolumn{2}{c}{Error/distance $\downarrow$} &
        \multicolumn{2}{c}{$r$ $\uparrow$} \\
      \cmidrule(lr){2-3}\cmidrule(l){4-5}
      Observable/metric & SupraTITO & DEP & SupraTITO & DEP \\
      \midrule
      SASA & $\mathbf{5.733(429)}$ & $15.167(125)$ &
        $\mathbf{.990(1)}$ & $.924(1)$ \\
      Contacts & $\mathbf{7.720(606)}$ & $18.802(231)$ &
        $\mathbf{.991(1)}$ & $.921(1)$ \\
      Cluster fraction & $\mathbf{.0343(27)}$ & $.0779(10)$ &
        $\mathbf{.982(1)}$ & $.917(2)$ \\
      Cluster $R_g$ & $\mathbf{.0405(36)}$ & $.1032(28)$ &
        $\mathbf{.981(2)}$ & $.897(5)$ \\
      Nematic order & $\mathbf{.0132(11)}$ & $.0311(10)$ &
        $\mathbf{.989(2)}$ & $.903(5)$ \\
      Shape anisotropy & $\mathbf{.0231(3)}$ & $.0404(6)$ &
        $\mathbf{.983(2)}$ & $.929(3)$ \\
      \addlinespace
      SASA $W_1$ & $\mathbf{6.03(32)}$ & $15.55(15)$ & -- & -- \\
      SASA KS & $\mathbf{.340(14)}$ & $.507(5)$ & -- & -- \\
      \bottomrule
    \end{tabular}
  \end{minipage}
  \caption{\textbf{Finite-lag learning versus direct ensemble prediction.}
  Both methods use the same training sequences and are evaluated on the same
  held-out set. \textbf{a}, sequence-resolved late-window SASA, ordered by the
  MD mean. Green, blue, and orange curves show MD, SupraTITO, and DEP,
  respectively. Generated curves show means across five independently sampled
  ensembles, and shaded bands show one standard deviation.
  \textbf{b}, mean absolute error and Pearson correlation for six structural
  observables, followed by the one-dimensional Wasserstein distance $W_1$ and
  Kolmogorov--Smirnov distance between generated and MD SASA distributions.
  Values in parentheses give the standard deviation across five stochastic
  generations in the final reported digits. SASA is reported in
  $\mathrm{nm}^2$, contact counts are dimensionless, and radius of gyration ($R_g$) is reported in
  nm.}
  \label{fig:direct-comparison}
\end{figure*}

SupraTITO follows the sequence-dependent MD SASA more closely than DEP,
which shows larger deviations for several sequences
(\cref{fig:direct-comparison}a). SupraTITO also achieves lower errors and higher
correlations with MD for all six structural observables
(\cref{fig:direct-comparison}b). Its SASA distributions have smaller
Wasserstein and Kolmogorov--Smirnov distances, showing closer agreement with
the corresponding MD ensembles. The coverage sweep provides a complementary
view of training-sequence efficiency: full-data DEP performs only slightly
better than SupraTITO trained on half as many sequences, whereas full-data
SupraTITO clearly outperforms DEP (\cref{fig:sequence-coverage}).

This advantage is consistent with how the two methods use the available
trajectories. DEP must infer the complete late assembly distribution
from sequence and molecular geometry alone. SupraTITO instead learns
configuration changes from many overlapping pairs along each trajectory, with
the current configuration providing information about the state of the
assembly. Repeated application then connects these learned changes into a
long trajectory. Similar benefits of finite-lag learning have been reported
for protein conformational dynamics~\citep{antoniadis2026platito}. SupraTITO
therefore produces more accurate late structures while also providing the
temporal evolution that is absent from direct ensemble prediction.

\subsection{SupraTITO transfers across peptide concentrations}
\label{sec:results-concentration}
We next test whether the learned dynamics respond to physical conditions outside
the training distribution. Peptide concentration is varied by changing the
periodic box volume while keeping the number of peptides fixed. The resulting
changes in intermolecular separation and periodic geometry are presented
directly to SupraTITO rather than encoded through a separate concentration
variable.

The benchmark contains 22 systems from six representative sequences selected
to span distinct concentration-dependent behavior. AG and KR remain largely
dispersed, FF and FH form stable aggregates, and IK exhibits a marked response
at low concentration. QI exhibits concentration-dependent assembly and further
tests concentration transfer for a sequence not seen during training. Complete simulation
conditions are provided in \cref{app:transfer-panel}.

\begin{figure*}[!htbp]
  \centering
  \includegraphics[width=0.96\textwidth]
  {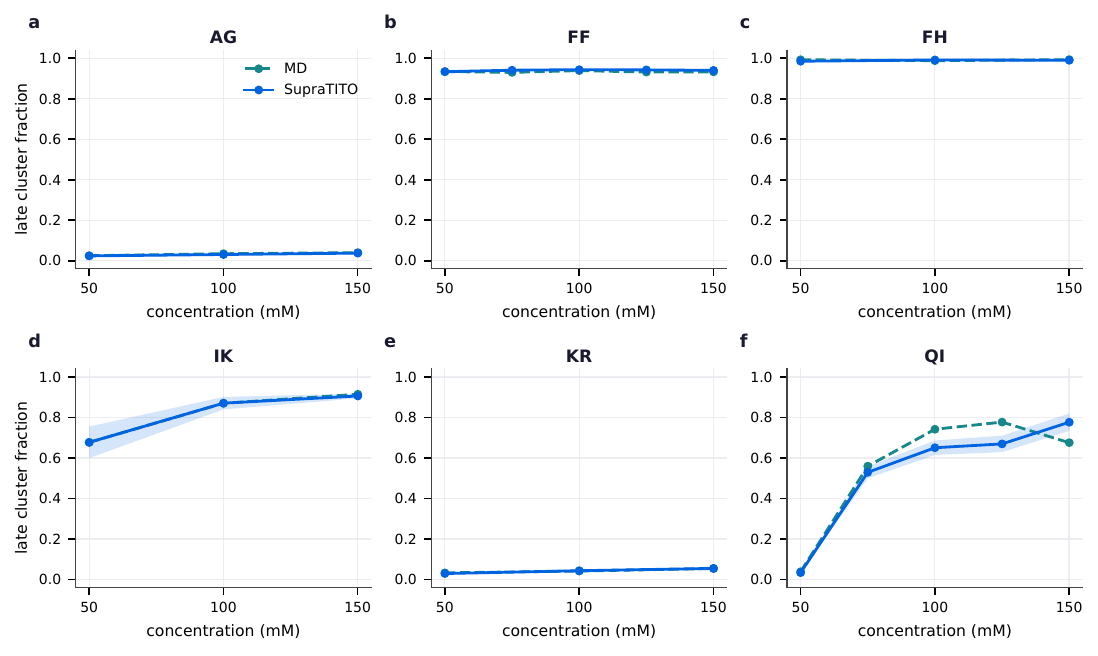}
  \caption{\textbf{Transfer beyond the training concentration.}
  Late largest-cluster fraction as a function of peptide concentration for six
  representative sequences at fixed peptide number. Dashed green curves show
  MD, solid blue curves show SupraTITO, and shaded regions show one standard
  deviation across generated rollouts. Complete SASA trajectories for each
  condition are provided in \cref{app:concentration-sasa}.}
  \label{fig:condition-transfer}
\end{figure*}

SupraTITO follows the MD concentration response across these systems, including
the most dilute conditions (\cref{fig:condition-transfer}). It preserves the
contrasting behavior of dispersed and strongly assembling peptides and
captures changes in the largest-cluster fraction as concentration varies. The
response of QI further shows that concentration transfer can be combined with
transfer to a held-out sequence.

The corresponding SASA trajectories show that this agreement is not limited to
the late assembly state (\cref{app:concentration-sasa}). SupraTITO reproduces the
progression of molecular exposure as well as its sequence-dependent late
values across the tested concentrations. These results suggest that the
learned operator responds to the intermolecular environments present in the
current configuration rather than assigning a fixed assembly outcome to each
sequence. Changes in concentration alter molecular separations and encounter
patterns, which are represented through the periodic pair geometry supplied
to the network.

We additionally vary the number of peptides to test execution beyond the
single system size used for training (\cref{app:size-transfer}). SupraTITO accepts
these variable-size configurations and reproduces some MD trends, but agreement
is less consistent than in the concentration study. This indicates that
architectural support for a variable number of molecular sites does not by
itself establish transfer across finite system sizes; training data spanning
multiple sizes may be required to learn the associated collective effects.

\FloatBarrier

\section{Conclusion}
\label{sec:conclusion}

We introduced SupraTITO, a transferable GenMD framework for collective
molecular dynamics. On a comprehensive dipeptide benchmark, it reproduces
assembly structures and dynamics for held-out sequences while preserving
molecular integrity over long rollouts. Predictions improve as training
sequences become more diverse, and finite-lag learning produces more accurate
late assembly structures than direct ensemble prediction trained on the same
trajectory collection. The learned dynamics also transfer across peptide
concentrations, including dilute conditions outside the training distribution.

These results show that a shared generative description can capture both
molecular integrity and collective assembly dynamics across peptide sequences
and concentrations. Beyond peptide assembly, the formulation provides a
foundation for modeling collective dynamics in other periodic supramolecular
systems.

\paragraph{Limitations and future work.}
The present study considers homogeneous dipeptide systems, with training
performed at a single concentration and system size. It therefore does
not yet establish transfer to longer peptides, co-assembly, or
arbitrary system sizes. A natural next step is to train across peptide lengths,
molecular compositions, concentrations, and box sizes, and to evaluate these
forms of generalization independently.

The reference data contain one MD trajectory for each condition. Peptide
assembly can follow multiple pathways and produce distinct metastable or
polymorphic morphologies, particularly in larger systems
\citep{wehner2020supramolecular}. Future benchmarks should therefore include
independent initializations and compare distributions of assembly times,
morphologies, and physical observables.

SupraTITO currently operates on a residue-level representation and reproduces the
dynamics generated by the Martini 2.1 force field. Its accuracy is consequently
limited by both the coarse-grained mapping and the reference simulator.
Atomistic descriptions are valuable when specific interactions, such as
hydrogen bonding and side-chain packing, determine assembly structure and
function. Future work could introduce atomistic detail directly through
architectures such as La-Proteina~\citep{geffner2026proteina}, or reconstruct
it using backmapping
~\citep{wassenaar2014going}. Backmapped configurations could then be
relaxed and propagated with a classical all-atom MD engine, combining SupraTITO's
efficient generation of long assembly trajectories with a higher-fidelity
description of local molecular interactions. Training on more accurate
reference simulations and validating the resulting dynamics against
experimental structural and kinetic observables would further improve their
physical reliability~\citep{smith2025calibrating}.

At the model level, flow matching directly on the periodic torus~\citep{chen2024flow} could provide
a more natural alternative to the current Euclidean interpolation of periodic
displacements. Global self-attention also scales quadratically with the number
of retained sites, which may become restrictive for longer peptides and larger
supramolecular systems. A scalable extension could combine local geometric
message passing with an efficient periodic long-range mechanism to preserve collective interactions without
evaluating every molecular pair explicitly \citep{kosmala2023ewald, frank2026machine}.

\section*{Acknowledgments}

We thank Simon Olsson and the members of the MFM Lab for helpful discussions.
We acknowledge the EuroHPC Joint Undertaking for awarding this project access
to the MareNostrum 5 supercomputer at the Barcelona Supercomputing Center
(BSC), Spain, under project ID EHPC-AIF-2025SC04-024. 
We also gratefully acknowledge the Gauss Centre for Supercomputing e.V. (www.gauss-centre.eu) for funding this project by providing computing time on the GCS Supercomputer SuperMUC-NG at Leibniz Supercomputing Centre (www.lrz.de).
This work was funded by the European Union through the ERC (StG SupraModel) - 101077842. Views and opinions expressed are however those of the author(s) only and do not necessarily reflect those of the European Union or the European Research Council Executive Agency. Neither the European Union nor the granting authority can be held responsible for them.

\section*{Data and Code Availability}

The data and code supporting this study will be made publicly available upon
acceptance.

\bibliographystyle{acl_natbib}
\bibliography{references}

@article{antoniadis2026platito,
  title={Protein Language Model Embeddings Improve Generalization of Implicit Transfer Operators},
  author={Antoniadis, Panagiotis and Pavesi, Beatrice and Olsson, Simon and Winther, Ole},
  journal={arXiv preprint arXiv:2602.11216},
  year={2026}
}

@article{lipman2022flow,
  title={Flow Matching for Generative Modeling},
  author={Lipman, Yaron and Chen, Ricky T. Q. and Ben-Hamu, Heli and Nickel, Maximilian and Le, Matt},
  journal={arXiv preprint arXiv:2210.02747},
  year={2022}
}

@inproceedings{albergo2023stochastic,
  title={Building Normalizing Flows with Stochastic Interpolants},
  author={Albergo, Michael S. and Vanden-Eijnden, Eric},
  booktitle={International Conference on Learning Representations},
  year={2023}
}

@article{noe2019boltzmann,
  title={Boltzmann Generators: Sampling Equilibrium States of Many-Body Systems with Deep Learning},
  author={No{\'e}, Frank and Olsson, Simon and K{\"o}hler, Jonas and Wu, Hao},
  journal={Science},
  volume={365},
  number={6457},
  pages={eaaw1147},
  year={2019}
}

@article{jumper2021highly,
  title={Highly accurate protein structure prediction with AlphaFold},
  author={Jumper, John and Evans, Richard and Pritzel, Alexander and Green, Tim and Figurnov, Michael and Ronneberger, Olaf and Tunyasuvunakool, Kathryn and Bates, Russ and {\v{Z}}{\'\i}dek, Augustin and Potapenko, Anna and others},
  journal={nature},
  volume={596},
  number={7873},
  pages={583--589},
  year={2021},
  publisher={Nature Publishing Group UK London}
}

@article{abramson2024accurate,
  title={Accurate structure prediction of biomolecular interactions with AlphaFold 3},
  author={Abramson, Josh and Adler, Jonas and Dunger, Jack and Evans, Richard and Green, Tim and Pritzel, Alexander and Ronneberger, Olaf and Willmore, Lindsay and Ballard, Andrew J and Bambrick, Joshua and others},
  journal={Nature},
  volume={630},
  number={8016},
  pages={493--500},
  year={2024},
  publisher={Nature Publishing Group UK London}
}

@article{baek2021accurate,
  title={Accurate prediction of protein structures and interactions using a three-track neural network},
  author={Baek, Minkyung and DiMaio, Frank and Anishchenko, Ivan and Dauparas, Justas and Ovchinnikov, Sergey and Lee, Gyu Rie and Wang, Jue and Cong, Qian and Kinch, Lisa N and Schaeffer, R Dustin and others},
  journal={Science},
  volume={373},
  number={6557},
  pages={871--876},
  year={2021},
  publisher={American Association for the Advancement of Science}
}

@article{lewis2025scalable,
  title={Scalable emulation of protein equilibrium ensembles with generative deep learning},
  author={Lewis, Sarah and Hempel, Tim and Jim{\'e}nez-Luna, Jos{\'e} and Gastegger, Michael and Xie, Yu and Foong, Andrew YK and Satorras, Victor Garc{\'\i}a and Abdin, Osama and Veeling, Bastiaan S and Zaporozhets, Iryna and others},
  journal={Science},
  volume={389},
  number={6761},
  pages={eadv9817},
  year={2025},
  publisher={American Association for the Advancement of Science}
}

@article{yang2026past,
  title={The past, present and future of de novo protein design},
  author={Yang, Wei and Wang, Shunzhi and Lee, Gyu Rie and Zhang, Jason Z and Courbet, Alexis and Juergens, David and Wang, Xinru and Schlichthaerle, Thomas and Abedi, Mohamad and Ragotte, Robert and others},
  journal={Nature},
  volume={652},
  number={8112},
  pages={1139--1152},
  year={2026},
  publisher={Nature Publishing Group UK London}
}

@article{reches2003casting,
  title={Casting metal nanowires within discrete self-assembled peptide nanotubes},
  author={Reches, Meital and Gazit, Ehud},
  journal={Science},
  volume={300},
  number={5619},
  pages={625--627},
  year={2003},
  publisher={American Association for the Advancement of Science}
}

@article{zheng2024predicting,
  title={Predicting equilibrium distributions for molecular systems with deep learning},
  author={Zheng, Shuxin and He, Jiyan and Liu, Chang and Shi, Yu and Lu, Ziheng and Feng, Weitao and Ju, Fusong and Wang, Jiaxi and Zhu, Jianwei and Min, Yaosen and others},
  journal={Nature Machine Intelligence},
  volume={6},
  number={5},
  pages={558--567},
  year={2024},
  publisher={Nature Publishing Group UK London}
}

@article{kaygisiz2025context,
  title={Context dependence in assembly code for supramolecular peptide materials and systems},
  author={Kaygisiz, K{\"u}bra and Sementa, Deborah and Athiyarath, Vignesh and Chen, Xi and Ulijn, Rein V},
  journal={Nature Reviews Materials},
  volume={10},
  number={6},
  pages={449--472},
  year={2025},
  publisher={Nature Publishing Group UK London}
}

@article{frederix2015exploring,
  title={Exploring the sequence space for (tri-) peptide self-assembly to design and discover new hydrogels},
  author={Frederix, Pim WJM and Scott, Gary G and Abul-Haija, Yousef M and Kalafatovic, Daniela and Pappas, Charalampos G and Javid, Nadeem and Hunt, Neil T and Ulijn, Rein V and Tuttle, Tell},
  journal={Nature chemistry},
  volume={7},
  number={1},
  pages={30--37},
  year={2015},
  publisher={Nature Publishing Group UK London}
}

@article{njirjak2024reshaping,
  title={Reshaping the discovery of self-assembling peptides with generative AI guided by hybrid deep learning},
  author={Njirjak, Marko and {\v{Z}}u{\v{z}}i{\'c}, Lucija and Babi{\'c}, Marko and Jankovi{\'c}, Patrizia and Otovi{\'c}, Erik and Kalafatovic, Daniela and Mau{\v{s}}a, Goran},
  journal={Nature machine intelligence},
  volume={6},
  number={12},
  pages={1487--1500},
  year={2024},
  publisher={Nature Publishing Group UK London}
}

@article{costa2026morphology,
  title={Morphology-Aware Peptide Discovery via Masked Conditional Generative Modeling},
  author={Costa, Nuno and Zavadlav, Julija},
  journal={ACS nano},
  volume={20},
  number={17},
  pages={13323},
  year={2026}
}

@article{shmilovich2020discovery,
  title={Discovery of self-assembling $\pi$-conjugated peptides by active learning-directed coarse-grained molecular simulation},
  author={Shmilovich, Kirill and Mansbach, Rachael A and Sidky, Hythem and Dunne, Olivia E and Panda, Sayak Subhra and Tovar, John D and Ferguson, Andrew L},
  journal={The Journal of Physical Chemistry B},
  volume={124},
  number={19},
  pages={3873--3891},
  year={2020},
  publisher={ACS Publications}
}

@article{olsson2026generative,
  title={Generative molecular dynamics},
  author={Olsson, Simon},
  journal={Current Opinion in Structural Biology},
  volume={96},
  pages={103213},
  year={2026},
  publisher={Elsevier}
}

@article{marrink2007martini,
  title={The MARTINI force field: coarse grained model for biomolecular simulations},
  author={Marrink, Siewert J and Risselada, H Jelger and Yefimov, Serge and Tieleman, D Peter and De Vries, Alex H},
  journal={The journal of physical chemistry B},
  volume={111},
  number={27},
  pages={7812--7824},
  year={2007},
  publisher={ACS Publications}
}

@article{schreiner2023implicit,
  title={Implicit transfer operator learning: Multiple time-resolution models for molecular dynamics},
  author={Schreiner, Mathias and Winther, Ole and Olsson, Simon},
  journal={Advances in Neural Information Processing Systems},
  volume={36},
  pages={36449--36462},
  year={2023}
}

@article{diez2026transferable,
  title={Transferable generative models bridge femtosecond to nanosecond time-step molecular dynamics},
  author={Diez, Juan Viguera and Schreiner, Mathias and Olsson, Simon},
  journal={Science advances},
  volume={12},
  number={15},
  pages={eaed2333},
  year={2026},
  publisher={American Association for the Advancement of Science}
}

@article{klein2023timewarp,
  title={Timewarp: Transferable acceleration of molecular dynamics by learning time-coarsened dynamics},
  author={Klein, Leon and Foong, Andrew and Fjelde, Tor and Mlodozeniec, Bruno and Brockschmidt, Marc and Nowozin, Sebastian and No{\'e}, Frank and Tomioka, Ryota},
  journal={Advances in Neural Information Processing Systems},
  volume={36},
  pages={52863--52883},
  year={2023}
}

@article{costa2024equijump,
  title={Equijump: Protein dynamics simulation via so (3)-equivariant stochastic interpolants},
  author={Costa, Allan Dos Santos and Mitnikov, Ilan and Pellegrini, Franco and Daigavane, Ameya and Geiger, Mario and Cao, Zhonglin and Kreis, Karsten and Smidt, Tess and Kucukbenli, Emine and Jacobson, Joseph},
  journal={arXiv preprint arXiv:2410.09667},
  year={2024}
}

@article{vlachas2022accelerated,
  title={Accelerated simulations of molecular systems through learning of effective dynamics},
  author={Vlachas, Pantelis R and Zavadlav, Julija and Praprotnik, Matej and Koumoutsakos, Petros},
  journal={Journal of Chemical Theory and Computation},
  volume={18},
  number={1},
  pages={538--549},
  year={2022},
  publisher={ACS Publications}
}

@misc{frenkel1997understanding,
  title={Understanding molecular simulation: from algorithms to applications},
  author={Frenkel, Daan and Smit, Berend and Ratner, Mark A},
  year={1997},
  publisher={American Institute of Physics}
}

@article{prinz2011markov,
  title={Markov models of molecular kinetics: Generation and validation},
  author={Prinz, Jan-Hendrik and Wu, Hao and Sarich, Marco and Keller, Bettina and Senne, Martin and Held, Martin and Chodera, John D and Sch{\"u}tte, Christof and No{\'e}, Frank},
  journal={The Journal of chemical physics},
  volume={134},
  number={17},
  year={2011},
  publisher={AIP Publishing}
}

@article{nuske2014variational,
  title={Variational approach to molecular kinetics},
  author={Nuske, Feliks and Keller, Bettina G and P{\'e}rez-Hern{\'a}ndez, Guillermo and Mey, Antonia SJS and No{\'e}, Frank},
  journal={Journal of chemical theory and computation},
  volume={10},
  number={4},
  pages={1739--1752},
  year={2014},
  publisher={ACS Publications}
}

@article{husic2018markov,
  title={Markov state models: From an art to a science},
  author={Husic, Brooke E and Pande, Vijay S},
  journal={Journal of the American Chemical Society},
  volume={140},
  number={7},
  pages={2386--2396},
  year={2018},
  publisher={ACS Publications}
}

@article{mori1965transport,
  title={Transport, collective motion, and Brownian motion},
  author={Mori, Hazime},
  journal={Progress of theoretical physics},
  volume={33},
  number={3},
  pages={423--455},
  year={1965},
  publisher={Oxford University Press}
}

@article{klippenstein2021introducing,
  title={Introducing memory in coarse-grained molecular simulations},
  author={Klippenstein, Viktor and Tripathy, Madhusmita and Jung, Gerhard and Schmid, Friederike and Van Der Vegt, Nico FA},
  journal={The journal of physical chemistry. B},
  volume={125},
  number={19},
  pages={4931},
  year={2021}
}

@article{liu2022flow,
  title={Flow straight and fast: Learning to generate and transfer data with rectified flow},
  author={Liu, Xingchao and Gong, Chengyue and Liu, Qiang},
  journal={arXiv preprint arXiv:2209.03003},
  year={2022}
}

@article{chen2018neural,
  title={Neural ordinary differential equations},
  author={Chen, Ricky TQ and Rubanova, Yulia and Bettencourt, Jesse and Duvenaud, David K},
  journal={Advances in neural information processing systems},
  volume={31},
  year={2018}
}

@article{he2026peptides,
  author  = {He, Z. and Bai, C. and Ma, R. and Zhang, S. and Fan, X. and Li, J. and Guo, S. and Luo, C. and Jin, J. and Ye, F.},
  title   = {Comprehensive Evaluation and Optimization of Martini Simulations for Dipeptide Self-Assembly},
  journal = {The Journal of Physical Chemistry B},
  year    = {2026},
  volume  = {130},
  pages   = {3358--3368},
  doi     = {10.1021/acs.jpcb.5c08159}
}

@article{peptidebuilder2013,
  author  = {Tien, Matthew Z. and Sydykova, Dariya K. and Meyer, Andrew G. and Wilke, Claus O.},
  title   = {PeptideBuilder: A Simple Python Library to Generate Model Peptides},
  journal = {PeerJ},
  year    = {2013},
  volume  = {1},
  pages   = {e80},
  doi     = {10.7717/peerj.80}
}

@article{abraham2015gromacs,
  author  = {Abraham, Mark James and Murtola, Teemu and Schulz, Roland and P{\'a}ll, Szil{\'a}rd and Smith, Jeremy C. and Hess, Berk and Lindahl, Erik},
  title   = {{GROMACS}: High Performance Molecular Simulations through Multi-Level Parallelism from Laptops to Supercomputers},
  journal = {SoftwareX},
  year    = {2015},
  volume  = {1--2},
  pages   = {19--25},
  doi     = {10.1016/j.softx.2015.06.001}
}

@article{jing2024generative,
  title={Generative modeling of molecular dynamics trajectories},
  author={Jing, Bowen and St{\"a}rk, Hannes and Jaakkola, Tommi and Berger, Bonnie},
  journal={Advances in Neural Information Processing Systems},
  volume={37},
  pages={40534--40564},
  year={2024}
}

@article{thiemann2026force,
  title={Force-free molecular dynamics through autoregressive equivariant networks},
  author={Thiemann, Fabian L and Resch{\"u}tzegger, Thiago and Esposito, Massimiliano and Taddese, Tseden and Olarte-Plata, Juan D and Martelli, Fausto},
  journal={Nature Machine Intelligence},
  pages={1--13},
  year={2026},
  publisher={Nature Publishing Group UK London}
}

@article{batra2022machine,
  title={Machine learning overcomes human bias in the discovery of self-assembling peptides},
  author={Batra, Rohit and Loeffler, Troy D and Chan, Henry and Srinivasan, Srilok and Cui, Honggang and Korendovych, Ivan V and Nanda, Vikas and Palmer, Liam C and Solomon, Lee A and Fry, H Christopher and others},
  journal={Nature chemistry},
  volume={14},
  number={12},
  pages={1427--1435},
  year={2022},
  publisher={Nature Publishing Group UK London}
}

@inproceedings{geffner2025proteina,
  title={Proteina: Scaling Flow-based Protein Structure Generative Models},
  author={Geffner, Tomas and Didi, Kieran and Zhang, Zuobai and Reidenbach, Danny and Cao, Zhonglin and Yim, Jason and Geiger, Mario and Dallago, Christian and Kucukbenli, Emine and Vahdat, Arash and Kreis, Karsten},
  booktitle={International Conference on Learning Representations},
  year={2025}
}

@article{dill2012protein,
  title={The protein-folding problem, 50 years on},
  author={Dill, Ken A and MacCallum, Justin L},
  journal={science},
  volume={338},
  number={6110},
  pages={1042--1046},
  year={2012},
  publisher={American Association for the Advancement of Science}
}

@article{henzler2007dynamic,
  title={Dynamic personalities of proteins},
  author={Henzler-Wildman, Katherine and Kern, Dorothee},
  journal={Nature},
  volume={450},
  number={7172},
  pages={964--972},
  year={2007},
  publisher={Nature Publishing Group UK London}
}

@article{lehn2002toward,
  title={Toward self-organization and complex matter},
  author={Lehn, Jean-Marie},
  journal={Science},
  volume={295},
  number={5564},
  pages={2400--2403},
  year={2002},
  publisher={American Association for the Advancement of Science}
}

@article{whitesides2002self,
  title={Self-assembly at all scales},
  author={Whitesides, George M and Grzybowski, Bartosz},
  journal={Science},
  volume={295},
  number={5564},
  pages={2418--2421},
  year={2002},
  publisher={American Association for the Advancement of Science}
}

@article{tong2023improving,
  title={Improving and generalizing flow-based generative models with minibatch optimal transport},
  author={Tong, Alexander and Fatras, Kilian and Malkin, Nikolay and Huguet, Guillaume and Zhang, Yanlei and Rector-Brooks, Jarrid and Wolf, Guy and Bengio, Yoshua},
  journal={arXiv preprint arXiv:2302.00482},
  year={2023}
}

@article{christians2026probabilistic,
  title={Probabilistic Forecasting for Coarse-Grained Molecular Dynamics},
  author={Christians, Luc F and Wojnar, Anna and Pak, Alexander J},
  journal={Journal of Chemical Theory and Computation},
  volume={22},
  number={9},
  pages={4231},
  year={2026}
}

@article{sidky2020molecular,
  title={Molecular latent space simulators},
  author={Sidky, Hythem and Chen, Wei and Ferguson, Andrew L},
  journal={Chemical Science},
  volume={11},
  number={35},
  pages={9459--9467},
  year={2020},
  publisher={The Royal Society of Chemistry}
}

@inproceedings{geffner2026proteina,
  title={La-proteina: Atomistic protein generation via partially latent flow matching},
  author={Geffner, Tomas and Didi, Kieran and Cao, Zhonglin and Reidenbach, Danny and Zhang, Zuobai and Dallago, Christian and Kucukbenli, Emine and Kreis, Karsten and Vahdat, Arash},
  booktitle={International Conference on Learning Representations},
  volume={2026},
  pages={67350--67395},
  year={2026}
}

@article{wassenaar2014going,
  title={Going backward: a flexible geometric approach to reverse transformation from coarse grained to atomistic models},
  author={Wassenaar, Tsjerk A and Pluhackova, Kristyna and Bockmann, Rainer A and Marrink, Siewert J and Tieleman, D Peter},
  journal={Journal of chemical theory and computation},
  volume={10},
  number={2},
  pages={676--690},
  year={2014},
  publisher={ACS Publications}
}

@article{wehner2020supramolecular,
  title={Supramolecular polymerization through kinetic pathway control and living chain growth},
  author={Wehner, Marius and W{\"u}rthner, Frank},
  journal={Nature Reviews Chemistry},
  volume={4},
  number={1},
  pages={38--53},
  year={2020},
  publisher={Nature Publishing Group UK London}
}

@article{smith2025calibrating,
  title={Calibrating generative models to distributional constraints},
  author={Smith, Henry D and Diamant, Nathaniel L and Trippe, Brian L},
  journal={arXiv preprint arXiv:2510.10020},
  year={2025}
}

@inproceedings{kosmala2023ewald,
  title={Ewald-based long-range message passing for molecular graphs},
  author={Kosmala, Arthur and Gasteiger, Johannes and Gao, Nicholas and G{\"u}nnemann, Stephan},
  booktitle={International Conference on Machine Learning},
  pages={17544--17563},
  year={2023},
  organization={PMLR}
}

@article{frank2026machine,
  title={Machine learning global atomic representations with euclidean fast attention},
  author={Frank, J Thorben and Chmiela, Stefan and M{\"u}ller, Klaus-Robert and Unke, Oliver T},
  journal={Nature Machine Intelligence},
  volume={8},
  number={3},
  pages={388--402},
  year={2026},
  publisher={Nature Publishing Group UK London}
}

@inproceedings{chen2024flow,
  title={Flow matching on general geometries},
  author={Chen, Ricky TQ and Lipman, Yaron},
  booktitle={International Conference on Learning Representations},
  volume={2024},
  pages={47922--47945},
  year={2024}
}

@article{chen2026coarse,
  title={Coarse-Grained Boltzmann Generators},
  author={Chen, Weilong and Zhao, Bojun and Eckwert, Jan and Zavadlav, Julija},
  journal={arXiv preprint arXiv:2602.10637},
  year={2026}
}

\beginappendix
\section{Molecular Dynamics Data and Evaluation Sets}

\subsection{Dipeptide MD Dataset}
\label{app:simulation-details}

The reference dataset contains one independently initialized CG-MD trajectory
for each of the 400 ordered dipeptides formed from the canonical amino acids.
Peptides are constructed with PeptideBuilder~\citep{peptidebuilder2013} and
mapped to the Martini 2.1 protein force field with neutral termini and a
$\beta$-sheet backbone assignment, following the screening parametrization of
He et al.~\citep{he2026peptides}. Each simulation contains 100 peptides of one
sequence in an initially cubic $11.840\,\mathrm{nm}$ periodic box, corresponding
to approximately $100\,\mathrm{mM}$. The systems are solvated with standard
Martini water, and Na$^+$ or Cl$^-$ counterions neutralize the side-chain
charge. No background salt is included.

Simulations are performed using GROMACS 2024.5~\citep{abraham2015gromacs}.
Following energy minimization, each system is equilibrated for $10\,\mathrm{ns}$
in the NPT ensemble at $303\,\mathrm{K}$ and $1\,\mathrm{bar}$ using a
$25\,\mathrm{fs}$ integration step, stochastic velocity-rescaling temperature
coupling, and Berendsen isotropic pressure coupling. Production continues for
$1.25\,\mu\mathrm{s}$ under the same integration step and thermodynamic
conditions. Electrostatics use reaction-field treatment with a
$1.1\,\mathrm{nm}$ cutoff for nonbonded interactions. Coordinates are stored
every $2.5\,\mathrm{ps}$, yielding 500,001 production configurations per
sequence. Peptide molecules are subsequently made whole across periodic
boundaries.

The stored trajectories contain all Martini peptide beads. SupraTITO retains the
backbone bead of each residue, giving two sites per peptide and $N=200$ sites
for each 100-peptide system.

\subsection{Training and Held-Out Sequences}
\label{app:splits}

The 400 ordered sequences are divided into 320 training sequences and 80
held-out sequences. For every heterodipeptide, the two reverse orderings (XY)
and (YX) are assigned to the same partition, preventing a held-out composition
from occurring in reverse order during training. All 20 homodipeptides remain
in the training set. The 40 held-out unordered pairs balance amino-acid
coverage and the distributions of late-time MD SASA and relative SASA change.
Both orderings of the pairs in \cref{tab:heldout-pairs} form the evaluation
set; all remaining ordered sequences form the training set.

\begin{table}[!htbp]
  \centering
  \small
  \caption{Held-out unordered dipeptide pairs. Both ordered versions of every
  listed pair are included in the evaluation set.}
  \label{tab:heldout-pairs}
  \begin{tabular}{@{}cccc@{}}
    \toprule
    \multicolumn{4}{c}{Held-out pair groups} \\
    \midrule
    AC & AE & AI & AT \\
    CH & CR & CT & DG \\
    DH & DN & DY & EQ \\
    ER & ES & FI & FL \\
    FV & FW & GK & GS \\
    GT & HK & HN & IQ \\
    IY & KM & KV & LQ \\
    LV & LY & MP & MT \\
    MW & NQ & NV & PR \\
    PS & PY & RW & SW \\
    \bottomrule
  \end{tabular}
\end{table}

\subsection{Sequence-Coverage Subsets}
\label{app:coverage-splits}

To measure how performance depends on sequence coverage, we construct three
nested training sets containing 40, 80, and 160 ordered sequences from the
full 320-sequence training partition. Reverse orderings remain grouped at
every coverage. The deterministic selection balances amino-acid frequencies
and the training-set distributions of late-time MD SASA and relative SASA
change. The 40-sequence set is contained in the 80-sequence set, which is
contained in the 160-sequence set. All four models use the same architecture,
optimization schedule, rollout protocol, and 80 held-out evaluation sequences.

\subsection{Condition and System-Size Evaluation Sets}
\label{app:transfer-panel}

We construct a further 28-system panel to evaluate transfer beyond the
concentration and system size used for training. It contains 22 concentration
conditions at fixed peptide count and six evaluations at new system sizes.
\Cref{tab:transfer-panel} summarizes the complete panel. Here,
\texttt{C} denotes concentration changes at fixed peptide count, \texttt{S}
denotes changes in system size at the baseline concentration, and \texttt{R}
denotes concentration changes obtained by varying peptide number at fixed box
size.

\begin{table}[!htbp]
  \centering
  \small
  \caption{
    Concentration and finite-size transfer panel.
    $M$ denotes the number of peptides and $L_0$ the initial cubic box length.
    The \texttt{C100} conditions are the corresponding baseline trajectories
    reused from the exhaustive corpus.
  }
  \label{tab:transfer-panel}
  \begin{tabular}{@{}>{\raggedright\arraybackslash}p{0.16\linewidth}>{\raggedright\arraybackslash}p{0.22\linewidth}>{\raggedright\arraybackslash}p{0.17\linewidth}>{\raggedright\arraybackslash}p{0.35\linewidth}@{}}
    \toprule
    Peptides & Condition IDs & Comparison & Conditions \\
    \midrule

    FF, QI &
    \texttt{C050}, \texttt{C075}, \texttt{C100}, \texttt{C125}, \texttt{C150} &
    Concentration &
    $M=100$, concentrations of $50$, $75$, $100$, $125$,
    and $150\,\mathrm{mM}$ obtained with
    $L_0=14.917$, $13.032$, $11.840$, $10.991$, and $10.343\,\mathrm{nm}$,
    respectively. \\[0.4em]

    FF, QI &
    \texttt{S050}, \texttt{C100}, \texttt{S200} &
    Finite size &
    $M=50$, $100$, and $200$ at $100\,\mathrm{mM}$, with
    $L_0=9.397$, $11.840$, and $14.917\,\mathrm{nm}$, respectively. \\[0.4em]

    FF &
    \texttt{R050}, \texttt{R150} &
    Fixed-box route &
    $M=50$ and $150$ in the fixed $L_0=11.840\,\mathrm{nm}$ box,
    corresponding to $50$ and $150\,\mathrm{mM}$. \\[0.4em]

    FH, IK, AG, KR &
    \texttt{C050}, \texttt{C100}, \texttt{C150} &
    Concentration &
    $M=100$ at $50$, $100$, and $150\,\mathrm{mM}$,
    with $L_0=14.917$, $11.840$, and $10.343\,\mathrm{nm}$,
    respectively. \\

    \bottomrule
  \end{tabular}
\end{table}

All additional systems follow the same simulation protocol as the exhaustive
corpus. Only peptide count, initial box size, and random insertion seed differ.

\FloatBarrier
\section{Implementation, Training, and Rollout Details}

\subsection{Architecture and Hyperparameters}
\label{app:architecture}

The reported SupraTITO model uses the translation-invariant, topology-aware, and
molecule-consistent formulation in \cref{sec:patito-architecture}. Its
condition and velocity networks are specified in
\cref{tab:architecture-details}. The amino-acid and chain-role embeddings have
widths 24 and 8, the physical-lag embedding has width 32, and the flow-time
embedding has width 128. The complete model contains approximately 3 million
parameters.

The flow-matching target is the molecule-consistent periodic displacement in
\cref{eq:patito-displacement}. The two sites of a peptide share one
minimum-image center displacement, while the change in their periodic bond
vector describes internal motion. Molecule consistency therefore changes the
target construction without introducing additional model parameters.

The flow-time embedding modulates the velocity-network pair representation
through adaptive layer normalization. Pair representations are constructed
once from the input geometry and topology during each network evaluation and
remain fixed across transformer layers. SupraTITO does not use triangle attention
or iterative pair updates.

\begin{table}[!htbp]
  \centering
  \caption{Architecture of the reported translation-invariant,
  molecule-consistent nonuniform-64 model.}
  \label{tab:architecture-details}
  \begin{tabular}{@{}lcc@{}}
    \toprule
    Component & Condition network & Velocity network \\
    \midrule
    Transformer layers & 4 & 6 \\
    Site-representation width & 64 & 128 \\
    Conditioning width & 64 & 128 \\
    Pair width & 32 & 64 \\
    Attention heads & 8 & 8 \\
    Distance categories & 64 & 64 \\
    Relative directions & Yes & Yes \\
    Directed topology categories & 3 & 3 \\
    Initial site representation & Zero vector & \(W_z\mathbf z_u\) \\
    Pair-value messages & Yes & Yes \\
    Pair conditioning & None & Flow time \\
    \bottomrule
  \end{tabular}
\end{table}

Training uses AdamW with learning rate \(10^{-4}\), zero weight decay, and
gradient-norm clipping at 0.1. Physical lags are sampled uniformly from the
integers 1 through 1000 saved frames, and sequences are sampled uniformly.
Each epoch contains 500,000 transition pairs. Both frames of a pair
receive the same random wrapped translation, and the Gaussian perturbation of
the flow path is \(\sigma=0.01\).

The model is trained for 1,000 epochs on NVIDIA H100 GPUs using bfloat16 mixed
precision and distributed data parallelism. The global batch size is 768,
giving 651 optimizer updates per epoch. The reported checkpoint is saved after
epoch 999 at optimization step 651,000.

The nonuniform-64 distance representation used for the main results has 63
ordered boundaries (in nm), compactly written as
\begin{equation}
\begin{split}
\{\beta_k\}_{k=1}^{63}={}&
\{0.100{:}0.025{:}0.600\}
\cup\{0.650{:}0.050{:}1.000\}\\
&\cup\{1.100{:}0.100{:}2.000\}
\cup\{2.200{:}0.200{:}4.000\}\\
&\cup\{4.250{:}0.250{:}6.000\}
\cup\{6.500,7.000,7.500,8.000,9.000,10.500\}.
\end{split}
\end{equation}
Here, \(a{:}h{:}b\) denotes the inclusive arithmetic sequence from \(a\) to
\(b\) with spacing \(h\). Distances below the first boundary and above the last
boundary occupy the first and final categories, respectively.

\subsection{Trajectory Generation}
\label{app:rollout-settings}

Reported SupraTITO trajectories use a lag of 1000 saved MD frames, corresponding
to \(2.5\,\mathrm{ns}\) per generated transition. The flow ODE for each
transition is integrated using 100 explicit Euler steps. Applying the operator
500 times produces a \(1.25\,\mu\mathrm{s}\) trajectory with 501 stored
configurations, including the initial state. Five trajectories are initialized
from the first frame of the corresponding MD trajectory and generated with
independent flow noise. Scalar late-time observables are averaged over the
final \(125\,\mathrm{ns}\). During inference, the repeatedly evaluated velocity
network is compiled using \texttt{torch.compile} in its default mode.

\subsection{Computational Cost}
\label{app:computational-cost}

Training the final model required approximately 1,300 H100 GPU-hours.
After compilation warm-up, one lag-1000 transition requires
$\approx 0.25\,\mathrm{s}$ for one system and $\approx 6.4\,\mathrm{s}$ for a
batch of 80 systems on one H100 GPU. A complete $1.25\,\mu\mathrm{s}$
single-system trajectory is estimated to require approximately
$2\,\mathrm{min}$ of inference compute.

\begin{table}[!htbp]
  \centering
  \caption{\textbf{Computational cost of reference simulation and SupraTITO.}
  Inference timings exclude compilation, file output, and post-processing.}
  \label{tab:computational-cost}
  \begin{tabular}{@{}lll@{}}
    \toprule
    Task & Hardware & Computational cost \\
    \midrule
    MD production, $1.25\,\mu\mathrm{s}$
      & Intel Xeon Platinum 8480+ and GPU Max 1550
      & $\approx 2\,\mathrm{h/system}$ \\
    SupraTITO training
      & 32 NVIDIA H100 GPUs
      & $\approx 1{,}300$ GPU-hours \\
    Lag-1000 transition, one system
      & One NVIDIA H100 GPU
      & $\approx 0.25\,\mathrm{s}$ \\
    Lag-1000 transition, 80 systems
      & One NVIDIA H100 GPU
      & $\approx 6.4\,\mathrm{s}$ \\
    \bottomrule
  \end{tabular}
\end{table}

\FloatBarrier
\section{Metric Definitions}
\label{app:metrics}

We evaluate solvent exposure, molecular integrity, aggregate structure, and
temporal correlations using observables computed from the retained coordinates.
All coordinate-derived quantities are evaluated frame by frame using the
orthorhombic minimum-image convention described in \cref{sec:patito-operator}. Unless
stated otherwise, distances are in nanometers, areas are in square nanometers,
and order parameters are dimensionless. We do not use a binary assembly label
as a primary metric because thresholding would discard structural and temporal
information.

\subsection{Evaluation Hierarchy and Notation}

Let a system contain $M$ dipeptides and $N=2M$ residue beads. For sequence
$s$, $\mathbf{x}^{\mathrm{MD}}_{s,t}$ denotes the available MD trajectory and
$\mathbf{x}^{\mathrm{Gen}}_{s,t}$ denotes the corresponding generated
trajectory. A coordinate-based observable is first evaluated independently for
every frame. Temporal summaries are then formed independently for each of five
generated trajectories. Their mean is the SupraTITO estimate and their sample
standard deviation describes variation across rollout seeds. The resulting
sequence-level MD and generated values are the paired observations used for
cross-sequence statistics.

This order matters. Frames from one trajectory are temporally correlated, and
peptides or bonds from the same frame are structurally correlated. Neither is
treated as an independent replicate when cross-sequence statistics are
computed. Autocorrelation is likewise evaluated independently within each
sequence before curves are summarized across sequences.

\begin{table}[!htbp]
  \centering
  \caption{Numerical parameters used for trajectory analysis unless stated
  otherwise in the corresponding figure caption.}
  \label{tab:metric-defaults}
  \begin{tabularx}{\linewidth}{@{}lclX@{}}
    \toprule
    Quantity & Symbol & Default & Role \\
    \midrule
    SASA bead radius & $r_b$ & $0.29\,\mathrm{nm}$ & Effective radius of one residue bead \\
    Probe radius & $r_p$ & $0.14\,\mathrm{nm}$ & Solvent-probe contribution \\
    Sphere directions & $P$ & 20 & Surface quadrature \\
    Contact cutoff & $r_c$ & $0.875\,\mathrm{nm}$ & Peptide-center contact graph \\
    Late interval & $T_\mathrm{late}$ & $125\,\mathrm{ns}$ & Physical-time averaging window \\
    Contact-memory interval & $T_\mathrm{pers}$ & $25\,\mathrm{ns}$ & Corrected persistence AUC \\
    Maximum ACF lag & $\ell_\mathrm{max}$ & $50\,\mathrm{ns}$ & Maximum displayed temporal offset \\
    \bottomrule
  \end{tabularx}
\end{table}

\subsection{Coarse-Grained Solvent Exposure}

The solvent-accessible surface area (SASA) used here is a PBC-aware,
Shrake--Rupley-style estimate computed from the representative residue beads.
It should not be interpreted as atomistic SASA. Each bead is assigned an
effective radius $r_b=0.29\,\mathrm{nm}$, and a solvent probe of radius
$r_p=0.14\,\mathrm{nm}$ gives the sampled surface radius
$r_s=r_b+r_p=0.43\,\mathrm{nm}$. For bead center $\mathbf{x}_i$, the surface
is represented by $P=20$ approximately uniform Fibonacci-sphere directions
$\mathbf{u}_p$:
\begin{equation}
  \mathbf{y}_{ip}=\mathbf{x}_i+r_s\mathbf{u}_p,
  \qquad p=1,\ldots,P.
\end{equation}
A surface point is accessible if it is not inside the sampled surface sphere
of any other bead. We denote this accessibility indicator by
\begin{equation}
  \chi_{ip}=\mathbb{I}\!\left[
  \min_{j\ne i}\left\|\operatorname{MIC}_{\pbcbox}
  (\mathbf{y}_{ip}-\mathbf{x}_j)\right\|_2\ge r_s\right].
\end{equation}
The total exposure of a frame is therefore
\begin{equation}
  \operatorname{SASA}(\mathbf{x})=
  \frac{4\pi r_s^2}{P}\sum_{i=1}^{N}\sum_{p=1}^{P}\chi_{ip}.
  \label{eq:sasa}
\end{equation}
Lower values indicate that a larger fraction of the bead surfaces is buried,
as generally occurs upon aggregation. The effective bead radius was selected
using training-reference frames only by minimizing RMSE to matched GROMACS
reference SASA computed from the original MD representation. A 20-direction
calibration gave an optimum near $0.2825\,\mathrm{nm}$; the radius was rounded and fixed at
$0.29\,\mathrm{nm}$. Generated and MD bead trajectories are evaluated with
identical radii, directions, and PBC handling.

\subsection{Peptide Centers and Contact Graph}

For peptide $q$, let $\mathbf{n}_q$ and $\mathbf{c}_q$ denote its N- and
C-terminal residue beads. The minimum-image terminal vector, wrapped peptide
center, and unit orientation are
\begin{equation}
  \mathbf{v}_q=\operatorname{MIC}_{\pbcbox}(\mathbf{c}_q-\mathbf{n}_q),
  \quad
  \mathbf{p}_q=\operatorname{wrap}_{\pbcbox}
  (\mathbf{n}_q+\tfrac12\mathbf{v}_q),
  \quad
  \widehat{\mathbf{v}}_q=\frac{\mathbf{v}_q}{\|\mathbf{v}_q\|_2}.
  \label{eq:peptide-center}
\end{equation}
This construction prevents a peptide crossing a periodic boundary from
acquiring an artificial center near the middle of the box.

We define an undirected peptide contact graph $G=(V,E)$ with one node per
peptide and adjacency matrix
\begin{equation}
  A_{qr}=\mathbb{I}\!\left[q\ne r\ \land\
  \left\|\operatorname{MIC}_{\pbcbox}
  (\mathbf{p}_q-\mathbf{p}_r)\right\|_2<r_c\right],
  \qquad r_c=0.875\,\mathrm{nm}.
  \label{eq:contact-graph}
\end{equation}
The framewise contact count is
\begin{equation}
  C=\sum_{q=1}^{M-1}\sum_{r=q+1}^{M}A_{qr},
  \label{eq:contact-count}
\end{equation}
so each peptide pair contributes at most one contact, irrespective of how its
two residue beads are arranged. The raw count is suitable for the present
fixed-$M$ benchmark; comparisons across different system sizes should instead
use a normalized contact density.

The cutoff was fixed from the first minimum following the first peak of the
training-MD peptide-center radial distribution function. Calibration used
$0.01\,\mathrm{nm}$ radial shells, shell-volume and instantaneous-box-volume
normalization, equal weighting of sequences, and Gaussian smoothing with
$0.025\,\mathrm{nm}$ width. Held-out trajectories were not used to select the
cutoff.

\subsection{Contact Memory}

Estimating long-time molecular diffusion requires unwrapped trajectories, or
equivalently the winding numbers that record successive crossings of the
periodic box. These are not retained by the processed trajectories, so we do
not report diffusion coefficients. Instead, baseline-corrected contact
persistence characterizes relative molecular dynamics directly from the
PBC-aware contact graph.

Contact persistence measures whether the same peptide pairs remain associated
over time, rather than only comparing the total number of contacts. For
contact adjacency $A_{qr}(n)$ at frame $n$, the intermittent contact
persistence at lag $\ell$ is
\begin{equation}
  P_{\mathrm{c}}(\ell)=
  \frac{
    \sum_{n}\sum_{q<r} A_{qr}(n)A_{qr}(n+\ell)
  }{
    \sum_{n}\sum_{q<r} A_{qr}(n)
  },
  \label{eq:contact-persistence}
\end{equation}
where the sums include time origins for which both frames lie within the
selected temporal window. Thus, $P_{\mathrm{c}}(\ell)$ is the probability that
a peptide pair in contact at one time is also in contact after lag $\ell$.
Because this is an intermittent definition, a contact may break and reform
during the interval.

Raw intermittent persistence approaches the probability of contact rather
than zero after pair identity has been forgotten. For a window containing
$N_w$ frames and $M$ peptides, we estimate this trajectory-specific baseline
as
\begin{equation}
  p_{\mathrm{c}}=
  \frac{1}{N_w\binom{M}{2}}
  \sum_{n=0}^{N_w-1}\sum_{q<r}A_{qr}(n),
  \label{eq:contact-occupancy}
\end{equation}
and define the corrected persistence
\begin{equation}
  \widetilde P_{\mathrm{c}}(\ell)=
  \frac{P_{\mathrm{c}}(\ell)-p_{\mathrm{c}}}
  {1-p_{\mathrm{c}}}.
  \label{eq:corrected-contact-persistence}
\end{equation}
The baseline is computed separately for the MD trajectory and for every
SupraTITO rollout before averaging across rollouts. Consequently,
$\widetilde P_{\mathrm{c}}(0)=1$ and the corrected curve approaches zero when
contact identity is lost. We summarize contact memory by
\begin{equation}
  \widetilde A_{\mathrm{c}}=
  \int_0^{25\,\mathrm{ns}}\widetilde P_{\mathrm{c}}(t)\,dt.
  \label{eq:contact-memory-auc}
\end{equation}
A larger $\widetilde A_{\mathrm{c}}$ indicates longer retention of peptide-pair
identity, not simply a larger number of contacts. Because the definition is
intermittent and generated configurations are separated by $2.5\,\mathrm{ns}$,
it is a contact-memory diagnostic rather than an uninterrupted contact lifetime
or a rate estimate below the output resolution.

\subsection{Largest-Cluster Size and Morphology}

Clusters are the connected components of the peptide contact graph. Let
$\mathcal{C}_\star$ be the component with the greatest number of peptides and
$M_\star=|\mathcal{C}_\star|$. Ties are resolved deterministically by the
lowest-indexed component. The largest-cluster fraction is
\begin{equation}
  f_\mathrm{LC}=\frac{M_\star}{M},
  \label{eq:largest-cluster-fraction}
\end{equation}
which ranges from $1/M$ for a fully disconnected frame to one when every
peptide belongs to a single connected component.

Cluster shape is evaluated from peptide centers rather than residue beads.
Because wrapped coordinates cannot be averaged directly, centers in the
largest cluster are first placed near an anchor $q_0\in\mathcal{C}_\star$,
\begin{equation}
  \widetilde{\mathbf{p}}_q=\mathbf{p}_{q_0}+
  \operatorname{MIC}_{\pbcbox}(\mathbf{p}_q-\mathbf{p}_{q_0}),
  \qquad q\in\mathcal{C}_\star.
  \label{eq:cluster-unwrap}
\end{equation}
With $\overline{\mathbf{p}}$ the mean unwrapped center, the gyration tensor is
\begin{equation}
  \mathbf{S}=\frac1{M_\star}\sum_{q\in\mathcal{C}_\star}
  (\widetilde{\mathbf{p}}_q-\overline{\mathbf{p}})
  (\widetilde{\mathbf{p}}_q-\overline{\mathbf{p}})^\top.
  \label{eq:gyration-tensor}
\end{equation}
For eigenvalues $0\le\lambda_1\le\lambda_2\le\lambda_3$, the cluster radius
of gyration and relative shape anisotropy are
\begin{equation}
  R_g=\sqrt{\lambda_1+\lambda_2+\lambda_3},
  \qquad
  \kappa^2=\frac{3}{2}
  \frac{\sum_{\alpha=1}^{3}
  (\lambda_\alpha-R_g^2/3)^2}{R_g^4}.
  \label{eq:shape-metrics}
\end{equation}
$R_g$ measures cluster extent and has units of nanometers. The dimensionless
$\kappa^2$ is zero for an isotropic eigenvalue spectrum and approaches one for
an ideal rod. For a singleton cluster, the implementation assigns $R_g=0$ and
$\kappa^2=0$; orientational order remains undefined. Anchor-based unwrapping
is reliable for compact, non-percolating clusters but can be ambiguous for a
cluster spanning more than half the box relative to its anchor. Shape metrics
for such frames should therefore be interpreted cautiously.

\subsection{Cluster Nematic Order}

Nematic order measures whether the molecular axes in the largest cluster share
a common orientation, while treating parallel and antiparallel axes as
equivalent. For an index set $\mathcal{I}$, define the traceless alignment
tensor
\begin{equation}
  \mathbf{Q}(\mathcal{I})=\frac{1}{|\mathcal{I}|}
  \sum_{q\in\mathcal{I}}
  \left(\frac32\widehat{\mathbf{v}}_q\widehat{\mathbf{v}}_q^\top
  -\frac12\mathbf{I}\right),
  \qquad
  S(\mathcal{I})=\lambda_\mathrm{max}\!\left(\mathbf{Q}(\mathcal{I})\right).
  \label{eq:nematic-order}
\end{equation}
Because $\mathbf{Q}$ is a real symmetric $3\times3$ matrix, it has three real
eigenvalues and mutually orthogonal eigenvectors. The largest eigenvalue
$\lambda_\mathrm{max}$ is the alignment strength $S$, and its eigenvector is
the nematic director: the spatial axis along which the peptide orientations
are most strongly aligned. The outer product
$\widehat{\mathbf{v}}_q\widehat{\mathbf{v}}_q^\top$ is unchanged when
$\widehat{\mathbf{v}}_q$ is replaced by $-\widehat{\mathbf{v}}_q$, so parallel
and antiparallel peptides contribute identically.

We evaluate \cref{eq:nematic-order} with
$\mathcal{I}=\mathcal{C}_\star$. Values near zero
indicate that no common orientation dominates, whereas values approaching one
indicate that nearly all molecular axes follow a common director. This
orientation metric is independent of whether the peptides are spatially close:
a compact aggregate can have low nematic order when its constituent peptides
are orientationally disordered. The metric is undefined for a singleton cluster
and is omitted from summaries for that frame.

\subsection{Intramolecular Bond Fidelity}

The indexed residue pair $(2q-1,2q)$ defines peptide $q$ throughout the MD and
generated trajectories. Its minimum-image intramolecular distance is
\begin{equation}
  d_q^\mathrm{bond}=
  \left\|\operatorname{MIC}_{\pbcbox}
  (\mathbf{c}_q-\mathbf{n}_q)\right\|_2.
  \label{eq:bond-distance}
\end{equation}
For a predeclared temporal window, bond lengths are pooled over frames and
peptide indices within each sequence and rollout.

\subsection{Temporal Summaries and Autocorrelation}

For a framewise observable $y_n$ with $n=0,\ldots,T-1$, we retain the complete
time series and report three scalar summaries: the trajectory mean, the final
value, and the mean over a fixed physical late interval. If output frames are
separated by $\Delta t$, the number of late frames is
\begin{equation}
  W=\min\!\left(T,
  \left\lceil\frac{T_\mathrm{late}}{\Delta t}\right\rceil\right),
  \qquad T_\mathrm{late}=125\,\mathrm{ns},
\end{equation}
and
\begin{equation}
  \overline{y}_\mathrm{late}=\frac1W
  \sum_{n=T-W}^{T-1}y_n.
  \label{eq:late-window}
\end{equation}
Using physical time rather than a fixed number of output frames makes the
summary comparable across rollout lags and output cadences. Summaries are
formed independently within each trajectory.

For a finite trajectory with mean $\bar{y}$ and variance
$s_y^2=T^{-1}\sum_{n=0}^{T-1}(y_n-\bar{y})^2$, the normalized autocorrelation
at output lag $\ell$ is
\begin{equation}
  \rho(\ell)=\frac{1}{(T-\ell)s_y^2}
  \sum_{n=0}^{T-\ell-1}
  (y_n-\bar{y})(y_{n+\ell}-\bar{y}).
  \label{eq:acf}
\end{equation}
We evaluate \cref{eq:acf} separately for SASA, contact count,
largest-cluster fraction, and cluster nematic order within the first and final
$250\,\mathrm{ns}$ of each trajectory, using temporal offsets up to
$50\,\mathrm{ns}$. Each generated rollout is evaluated separately before its
autocorrelation is summarized across rollout seeds. Self-assembly
trajectories are generally nonstationary, so \(\rho(\ell)\) is interpreted as
a descriptive comparison of temporal persistence rather than as an
equilibrium correlation function or an independent kinetic-rate estimate.

\subsection{Across-Sequence Agreement}

Let $a_s$ and $\widehat{a}_s$ be the MD and generated summaries for
observable $a$ on sequence
$s\in\{1,\ldots,N_{\mathrm{seq}}\}$. We report the mean absolute
error
\begin{equation}
  \operatorname{MAE}
  =
  \frac1{N_{\mathrm{seq}}}\sum_{s=1}^{N_{\mathrm{seq}}}
  |\widehat{a}_s-a_s|.
  \label{eq:sequence-errors}
\end{equation}
Pearson correlation measures linear agreement across sequences,
\begin{equation}
  r=\frac{\sum_s(a_s-\bar{a})(\widehat{a}_s-\overline{\widehat{a}})}
  {\sqrt{\sum_s(a_s-\bar{a})^2}
   \sqrt{\sum_s(\widehat{a}_s-\overline{\widehat{a}})^2}},
  \label{eq:pearson}
\end{equation}
where \(\bar a\) and \(\overline{\widehat a}\) are the corresponding
across-sequence means. For comparisons between observables with different
units, \cref{fig:sequence-coverage} reports the range-normalized MAE
\begin{equation}
  \operatorname{nMAE}
  =
  \frac{\operatorname{MAE}}
  {\max_s a_s-\min_s a_s}.
  \label{eq:nmae}
\end{equation}
This normalization is used only for the visual comparison across observables;
\cref{fig:direct-comparison}b reports each MAE in its physical units.

For distributional comparisons, let \(F_s^{\mathrm{MD}}\) and
\(F_s^{\mathrm{Gen}}\) be the empirical cumulative distributions of an
observable for sequence \(s\). We use the one-dimensional Wasserstein and
Kolmogorov--Smirnov distances
\begin{equation}
  W_{1,s}
  =
  \int_{-\infty}^{\infty}
  \left|F_s^{\mathrm{Gen}}(a)-F_s^{\mathrm{MD}}(a)\right|\,da,
  \qquad
  D_{\mathrm{KS},s}
  =
  \sup_a
  \left|F_s^{\mathrm{Gen}}(a)-F_s^{\mathrm{MD}}(a)\right|.
  \label{eq:distribution-errors}
\end{equation}
The distributional values in \cref{fig:direct-comparison}b are arithmetic
means of these
per-sequence distances. Wasserstein distance retains the units of the
observable, whereas the Kolmogorov--Smirnov distance is dimensionless.

\section{Complete Results for Held-Out Sequences}
\label{app:heldout-complete}

The main text summarizes agreement across sequences and highlights three
representative trajectories. Here we report every held-out sequence to expose
the full range of outcomes. All figures use the SupraTITO checkpoint and lag-1000
protocol in \cref{app:rollout-settings}. Unless noted otherwise, generated
curves and markers show the mean across five rollouts, and bands and bars show
one standard deviation. Static configurations are taken from one
representative rollout because averaging coordinates across trajectories does
not produce a physical configuration.

\subsection{Final-Frame Agreement Across Sequences}
\label{app:heldout-final-scatter}

\clearpage
\subsection{Temporal Trajectories}
\label{app:heldout-temporal}

\begin{figure*}[!htbp]
  \centering
  \includegraphics[width=0.96\textwidth]{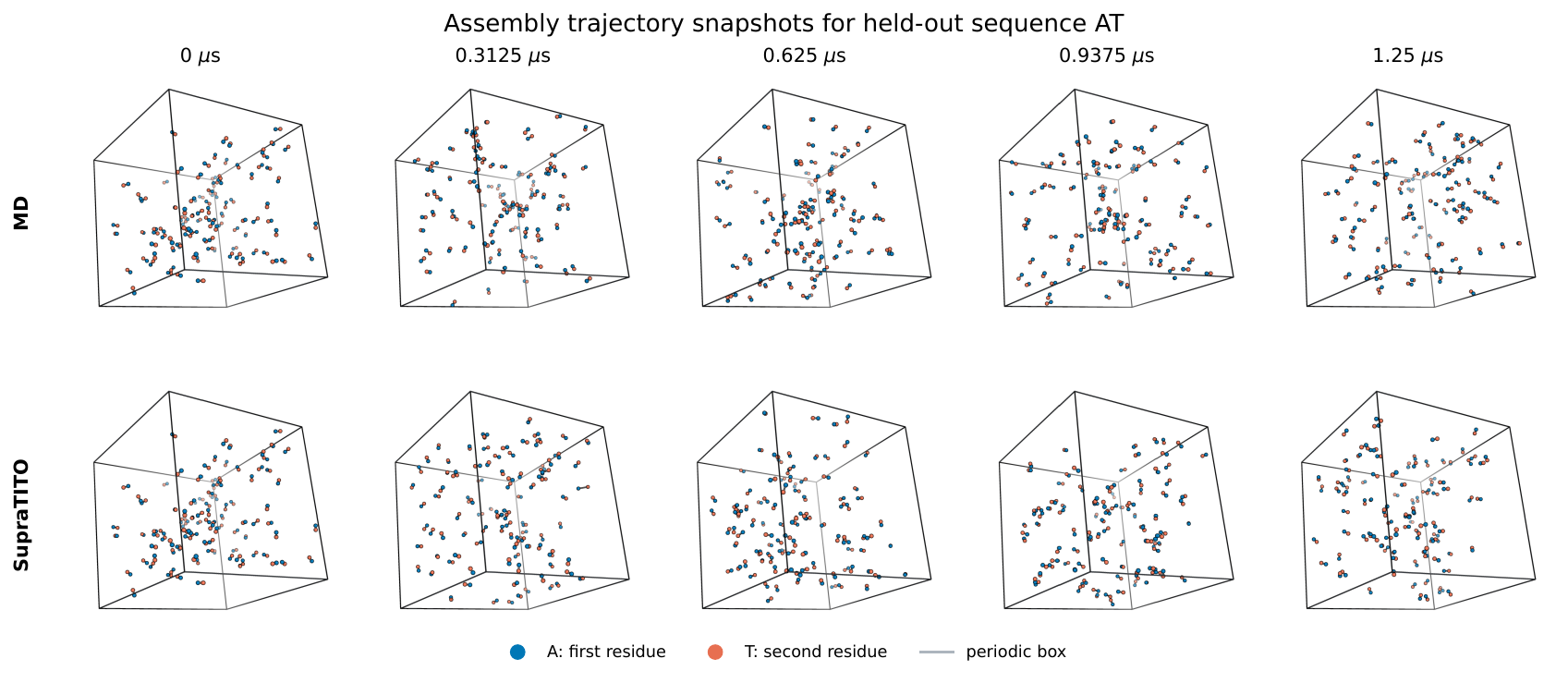}
  \caption{\textbf{Matched assembly snapshots for the representative
  dispersed sequence AT.} MD configurations are shown above the corresponding
  SupraTITO configurations at five physical times. Both trajectories retain a
  predominantly dispersed peptide population throughout the rollout. The
  first and second sequence positions are colored blue and coral,
  respectively.}
  \label{fig:trajectory-snapshots-at}
\end{figure*}

\begin{figure*}[!htbp]
  \centering
  \includegraphics[width=0.96\textwidth]{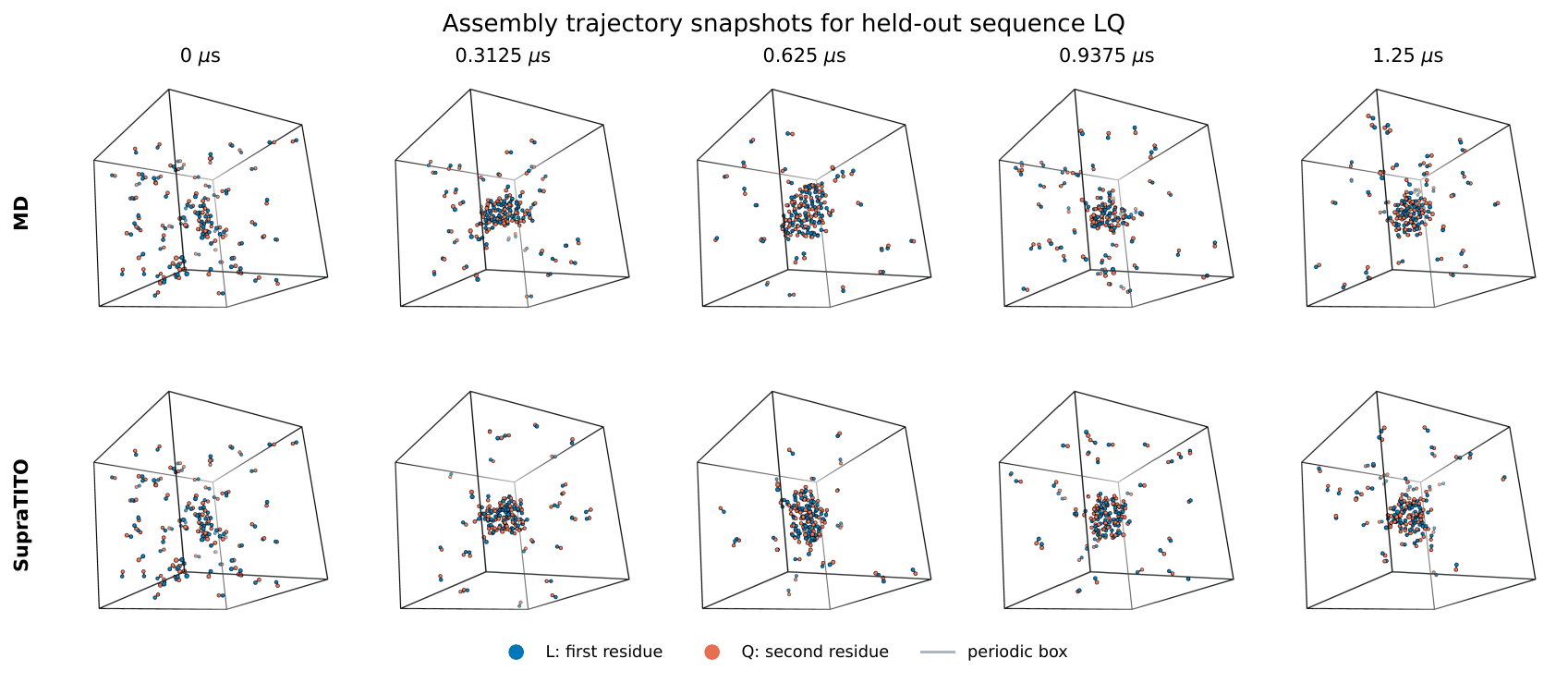}
  \caption{\textbf{Matched assembly snapshots for the representative
  intermediate sequence LQ.} MD and SupraTITO both form a substantial cluster
  while retaining a population of unassociated peptides, consistent with the
  partial and fluctuating assembly dynamics in \cref{fig:dynamics-bonds}b.}
  \label{fig:trajectory-snapshots-lq}
\end{figure*}

\begin{figure*}[!htbp]
  \centering
  \includegraphics[width=0.96\textwidth]{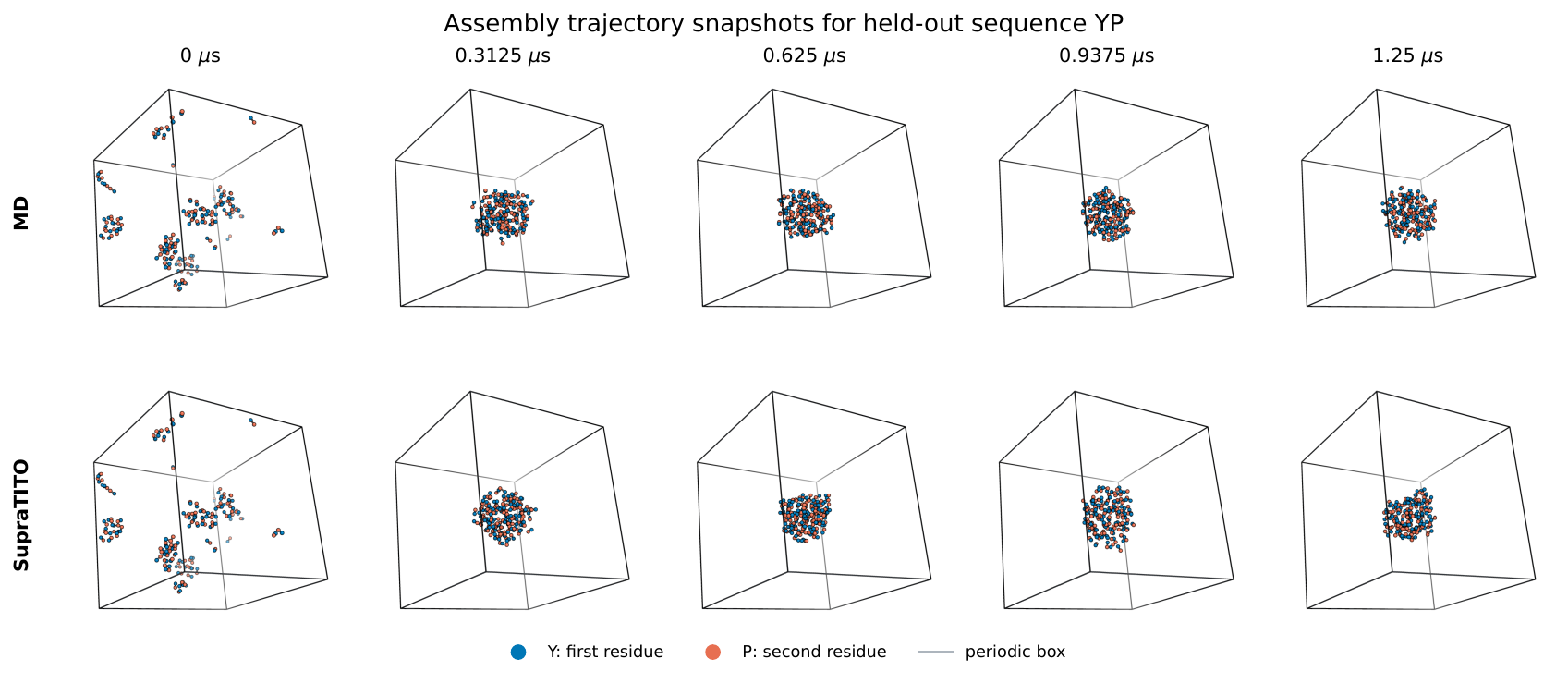}
  \caption{\textbf{Matched assembly snapshots for the representative held-out
  assembled sequence YP.} The two stochastic trajectories follow different
  microscopic pathways but both rapidly form a compact aggregate, consistent with
  \cref{fig:dynamics-bonds}c.}
  \label{fig:trajectory-snapshots-yp}
\end{figure*}

To complement the scalar trajectories,
\cref{fig:trajectory-snapshots-at,fig:trajectory-snapshots-lq,fig:trajectory-snapshots-yp}
show matched three-dimensional configurations for the dispersed,
intermediate, and assembled examples in the main text. Each sphere is one
retained residue site, gray sticks connect the two indexed sites of each
peptide, and the wireframe marks the periodic box. All panels use the same
camera orientation and box scale. For visualization, the largest
PBC-connected peptide cluster is unwrapped and centered, and the remaining
peptides are reimaged around it. This operation changes only the displayed
periodic images and preserves all minimum-image distances.

\FloatBarrier
\subsection{Final Configurations and Molecular Connectivity}
\label{app:heldout-xy}

\begin{figure*}[!htbp]
  \centering
  \includegraphics[width=0.92\textwidth,height=0.82\textheight,keepaspectratio]{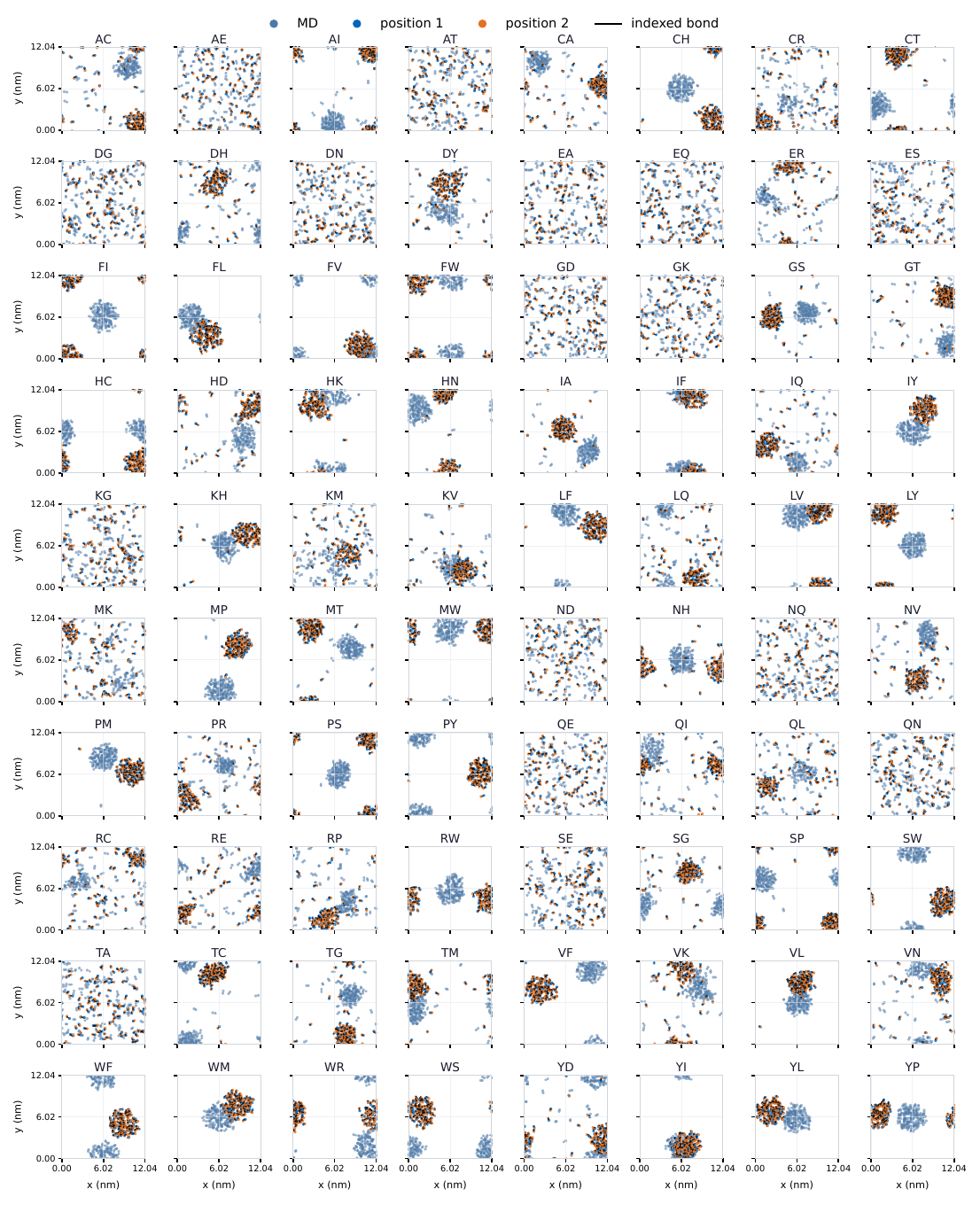}
  \caption{\textbf{Final configurations for all held-out sequences.}
  Muted blue points show the MD reference. TUM blue and orange points show the
  first and second residue positions in SupraTITO, respectively, and black lines
  connect the indexed sites of each peptide using minimum-image geometry.
  Generated coordinates are from one representative rollout. The XY
  projection supports visual comparison of morphology and molecular
  connectivity but can hide separation along the third coordinate.}
  \label{fig:all-unseen-xy}
\end{figure*}

\Cref{fig:all-unseen-xy} compares the final MD and generated morphologies and
shows the indexed intramolecular connectivity of the SupraTITO configurations for
every held-out sequence.

\FloatBarrier
\subsection{Intramolecular Bond Fidelity}
\label{app:heldout-bonds}

\begin{figure*}[!htbp]
  \centering
  \includegraphics[width=0.92\textwidth,height=0.82\textheight,keepaspectratio]{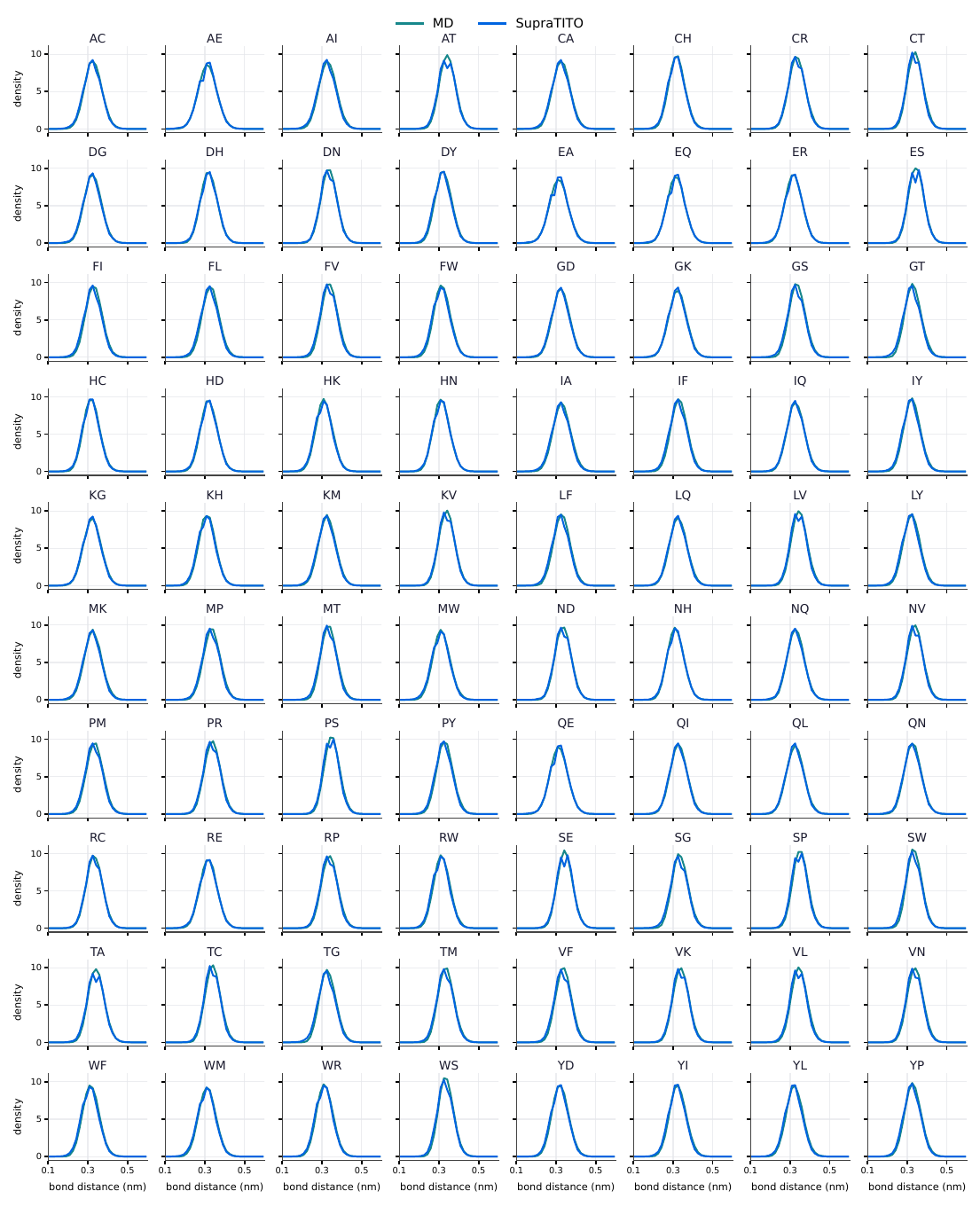}
  \caption{\textbf{Intramolecular bond-distance distributions for all held-out
  sequences.} Distances are computed for fixed indexed residue pairs with the
  full three-dimensional minimum-image convention and pooled over analyzed
  frames and peptides within each rollout. Each panel compares the MD
  distribution with the SupraTITO mean and one standard deviation across
  rollouts.}
  \label{fig:all-unseen-bonds}
\end{figure*}

The complete bond-distance comparison in \cref{fig:all-unseen-bonds} tests
whether the agreement in collective assembly is accompanied by physically
consistent intramolecular separations for every held-out sequence.

\FloatBarrier
\section{Direct Ensemble Prediction Baseline}
\label{app:direct-supplement}

The direct ensemble prediction (DEP) baseline learns the
sequence-conditioned late-structure density
\begin{equation}
  p_\phi(\mathbf{x}_\mathrm{late}\mid
  \sequence,\mathcal{G},\pbcbox)
\end{equation}
with flow matching along a linear probability path. Unlike SupraTITO, DEP
receives neither a current configuration nor a physical lag. Its periodic
Gaussian source draws each residue coordinate from an isotropic Gaussian with
standard deviation $1\,\mathrm{nm}$ around a uniformly sampled box origin.
Because peptides of the same sequence are exchangeable, source and target
peptide centers are matched by linear-sum assignment. The flow interpolates
minimum-image peptide-center displacements and center-relative residue
offsets, keeping each indexed peptide intact along the periodic path.

Targets are sampled uniformly from the final 50,000 saved frames,
corresponding to the final $125\,\mathrm{ns}$ of each training trajectory.
This window supplies multiple configurations per sequence and retains
structural variation within the late ensemble; using only one endpoint per
sequence would leave too few targets for conditional flow matching. DEP and
SupraTITO therefore use the same MD trajectory collection, with DEP trained on
the late ensemble and SupraTITO trained on finite-lag transitions throughout the
trajectories.

The baseline uses a single seven-layer Proteína transformer with site width 128,
pair width 64, and eight attention heads. Its inputs comprise flow time,
amino-acid identity, terminal role, directed bond topology, and minimum-image
pair-distance categories. It is trained for 1,000 epochs with 500,000 sampled
examples per epoch, AdamW at learning rate $10^{-4}$, and flow noise
$\sigma=0.01$. For each held-out sequence, the matched evaluation uses the
same 50 lag-strided frames from the MD late window as the transfer analysis,
250 independent DEP samples (50 from each of
five inference seeds), and 250 configurations sampled evenly from the final
$125\,\mathrm{ns}$ of five lag-1000 SupraTITO rollouts. Agreement metrics are
computed separately for each generated ensemble and reported as the mean and
sample standard deviation across the five ensembles.
The independent DEP samples support structural-distribution comparisons but
do not define a trajectory or an autocorrelation function.

\begin{figure*}[!htbp]
  \centering
  \includegraphics[width=0.92\textwidth]{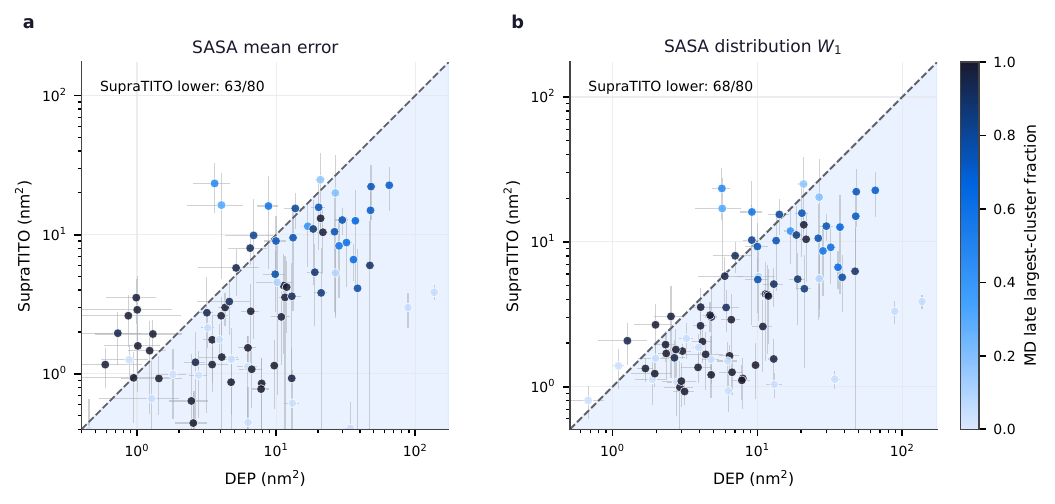}
  \caption{\textbf{Sequence-wise SASA errors for DEP and
  SupraTITO.} Each point is the mean for one held-out sequence, with horizontal
  and vertical bars showing one standard deviation across five DEP generation
  seeds and five SupraTITO rollout seeds, respectively. Points below the
  equal-error diagonal favor SupraTITO. Color denotes the MD late-window largest-cluster
  fraction, and logarithmic axes retain both low-error systems and outliers.}
  \label{fig:direct-sequence-errors}
\end{figure*}

\begin{figure*}[!htbp]
  \centering
  \includegraphics[width=0.92\textwidth,height=0.82\textheight,keepaspectratio]{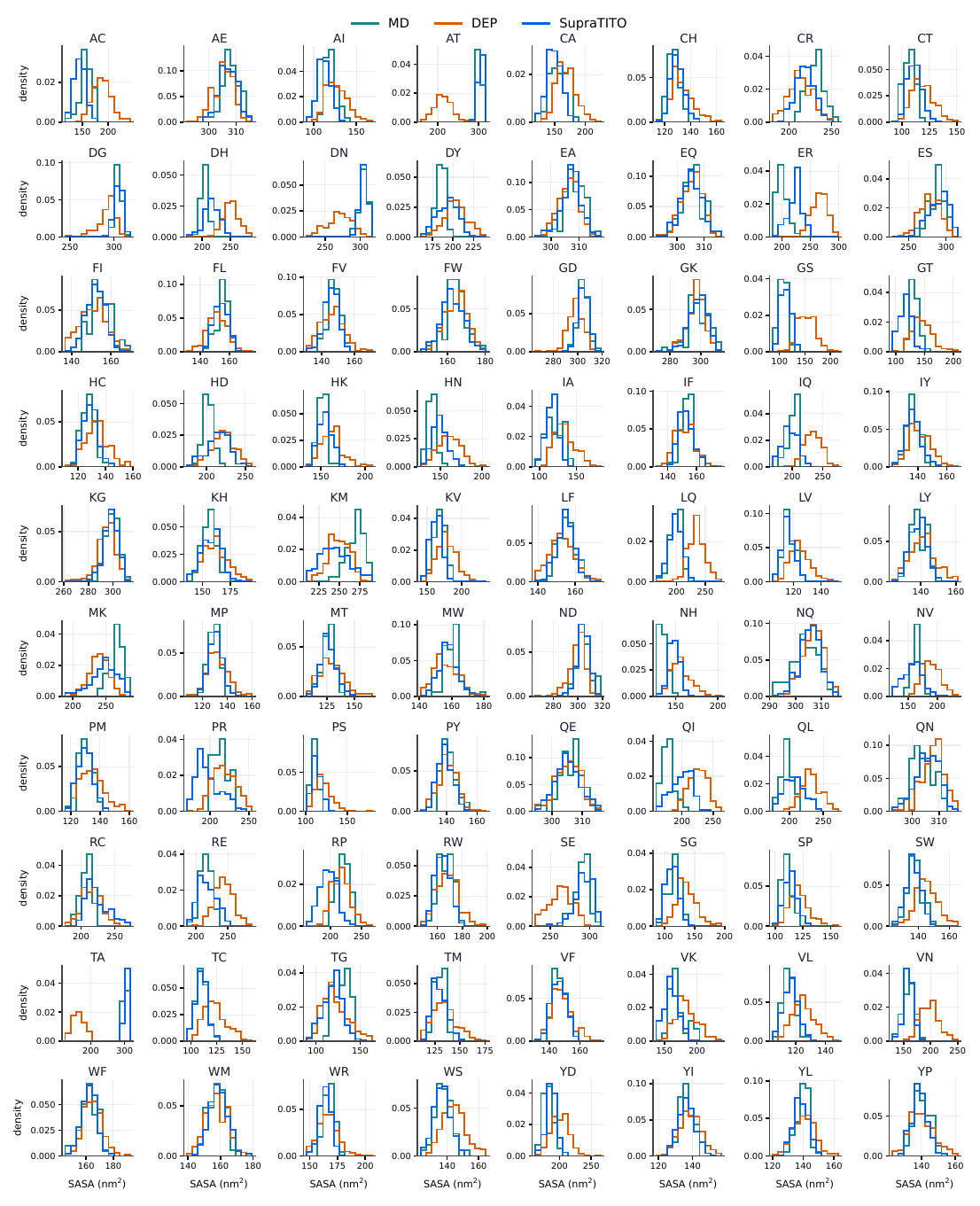}
  \caption{\textbf{Complete sequence-resolved SASA distributions for the
  DEP comparison.} Each panel contains 50 MD samples, 250 independent DEP
  predictions, and 250 SupraTITO late-window configurations. Per-panel
  axes retain the sequence-specific SASA range and probability-density scale.}
  \label{fig:direct-sasa-distributions}
\end{figure*}

The aggregate comparison in \cref{fig:direct-comparison} is complemented by
two sequence-resolved views. \Cref{fig:direct-sequence-errors} shows how the
SASA error is distributed across the held-out set, while
\cref{fig:direct-sasa-distributions} compares the complete generated and MD
SASA distributions for every sequence.

\FloatBarrier
\section{Condition and System-Size Transfer}
\label{app:size-transfer}

\subsection{Concentration-Resolved Exposure Dynamics}
\label{app:concentration-sasa}

The main text summarizes concentration transfer using the late
largest-cluster fraction. Here, each concentration is shown separately so that
solvent exposure and assembly timing can be compared without overlapping
curves. SASA is normalized by peptide count. SupraTITO curves summarize five
independent lag-1000 rollouts from the same initial MD configuration.

\begin{figure*}[!htbp]
  \centering
  \includegraphics[width=0.92\textwidth]{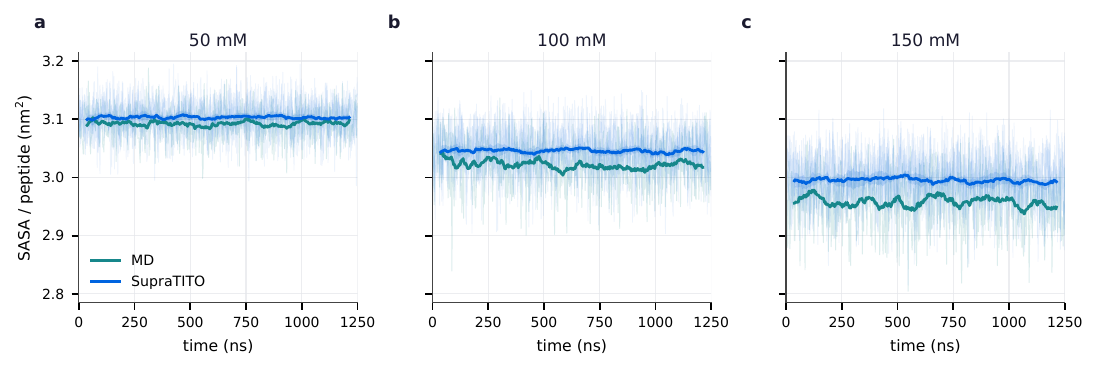}
  \caption{\textbf{Concentration-resolved SASA for AG.} Each panel is one
  100-peptide system at the indicated concentration. The green lines show the
  raw MD values and their 31-frame moving average. Faint blue lines show the
  individual rollouts; the solid blue line and shaded band show the mean and
  one standard deviation of their 31-frame moving averages.}
  \label{fig:concentration-sasa-ag}
\end{figure*}

\begin{figure*}[!htbp]
  \centering
  \includegraphics[width=0.92\textwidth]{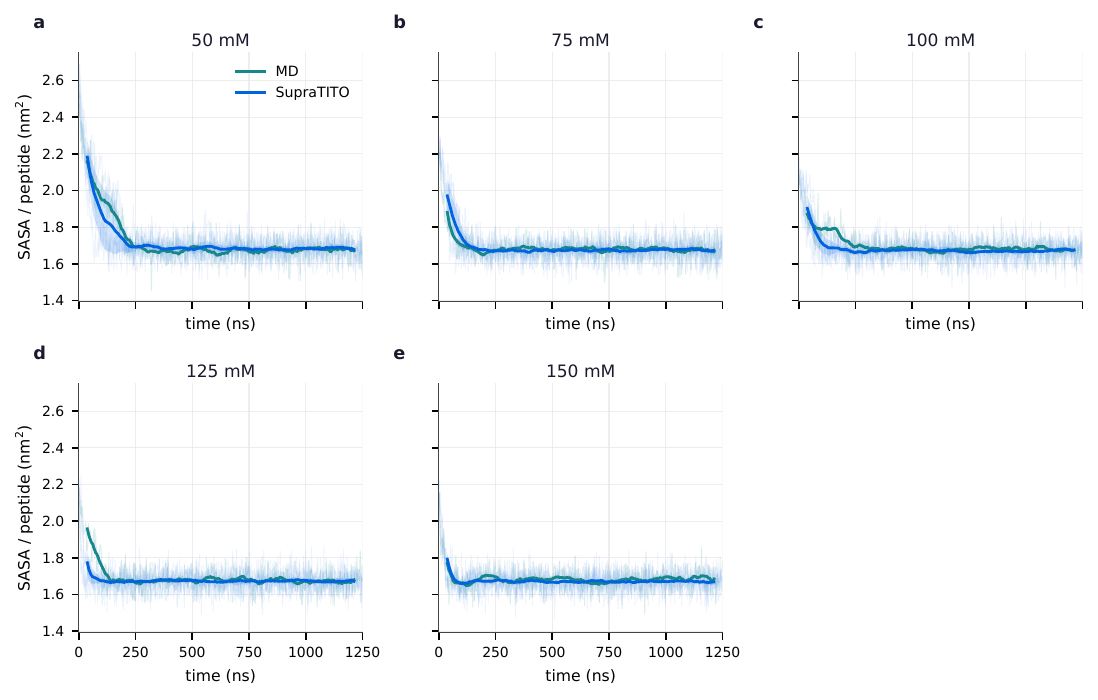}
  \caption{\textbf{Concentration-resolved SASA for FF.} Plotting conventions
  follow \cref{fig:concentration-sasa-ag}.}
  \label{fig:concentration-sasa-ff}
\end{figure*}

\begin{figure*}[!htbp]
  \centering
  \includegraphics[width=0.92\textwidth]{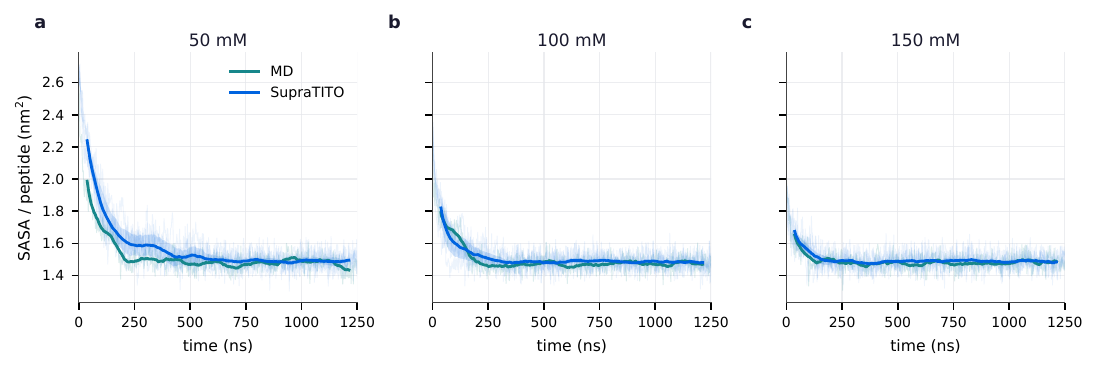}
  \caption{\textbf{Concentration-resolved SASA for FH.} Plotting conventions
  follow \cref{fig:concentration-sasa-ag}.}
  \label{fig:concentration-sasa-fh}
\end{figure*}

\begin{figure*}[!htbp]
  \centering
  \includegraphics[width=0.92\textwidth]{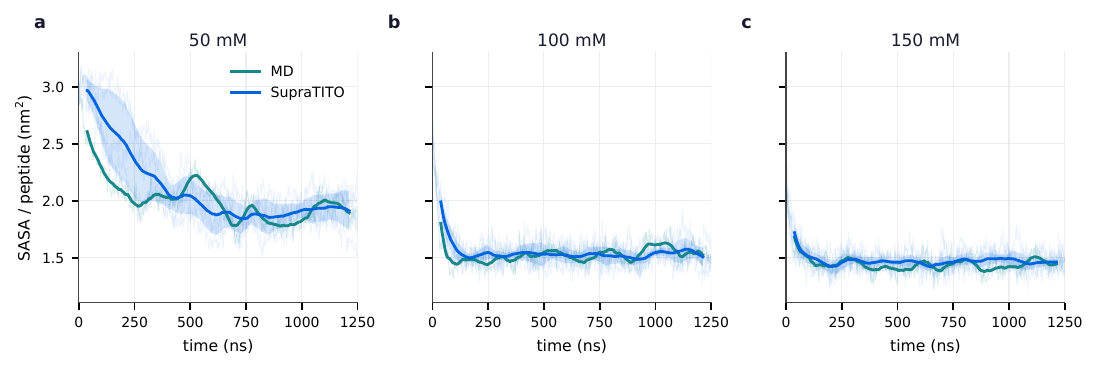}
  \caption{\textbf{Concentration-resolved SASA for IK.} Plotting conventions
  follow \cref{fig:concentration-sasa-ag}.}
  \label{fig:concentration-sasa-ik}
\end{figure*}

\begin{figure*}[!htbp]
  \centering
  \includegraphics[width=0.92\textwidth]{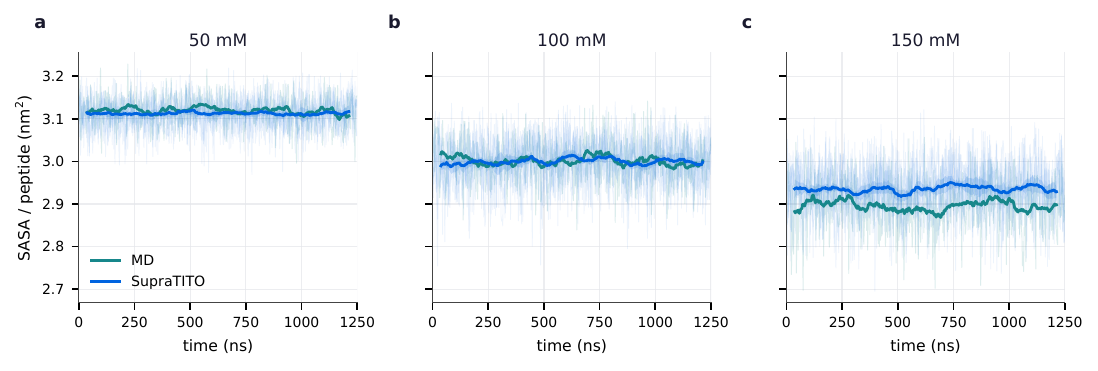}
  \caption{\textbf{Concentration-resolved SASA for KR.} Plotting conventions
  follow \cref{fig:concentration-sasa-ag}.}
  \label{fig:concentration-sasa-kr}
\end{figure*}

\begin{figure*}[!htbp]
  \centering
  \includegraphics[width=0.92\textwidth]{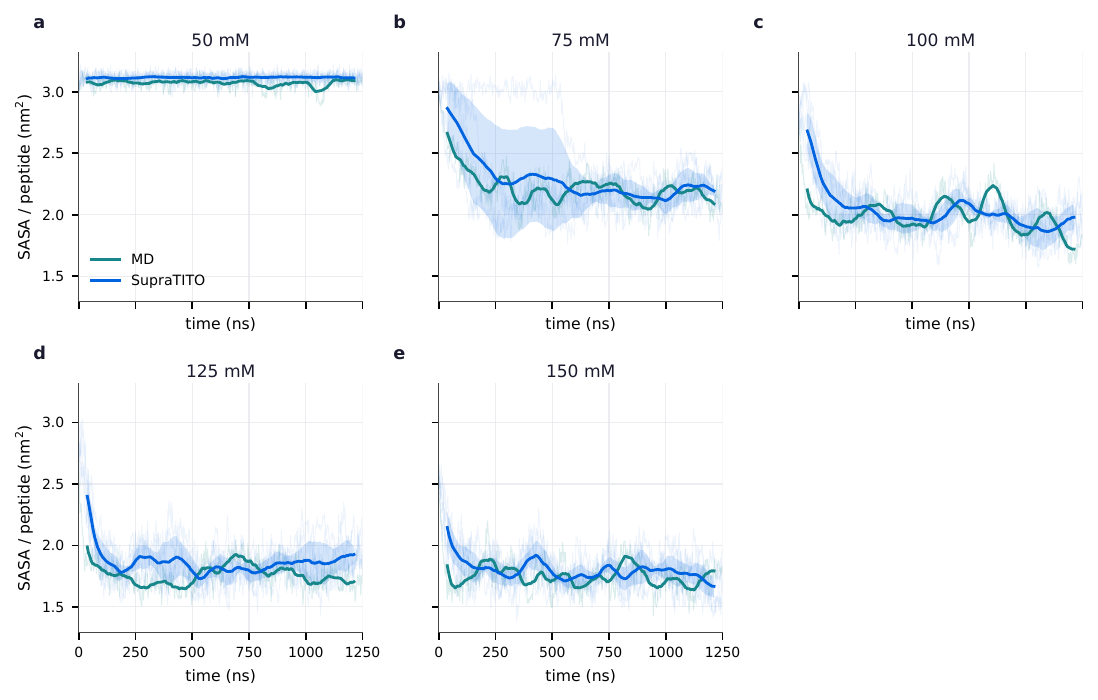}
  \caption{\textbf{Concentration-resolved SASA for QI.} Plotting conventions
  follow \cref{fig:concentration-sasa-ag}.}
  \label{fig:concentration-sasa-qi}
\end{figure*}

Across the 22 fixed-peptide-count conditions, the SASA trajectories in
\cref{fig:concentration-sasa-ag,fig:concentration-sasa-ff,fig:concentration-sasa-fh,fig:concentration-sasa-ik,fig:concentration-sasa-kr,fig:concentration-sasa-qi}
support the late cluster-fraction trends in \cref{fig:condition-transfer}.
SupraTITO preserves the distinct responses of dispersed, assembled, and
concentration-sensitive sequences, including at the most dilute conditions.
For QI, the concentration response is recovered even though the peptide
sequence is held out from training.

\FloatBarrier
\subsection{Evaluation at New System Sizes}

\begin{figure*}[!htbp]
  \centering
  \includegraphics[width=0.92\textwidth]{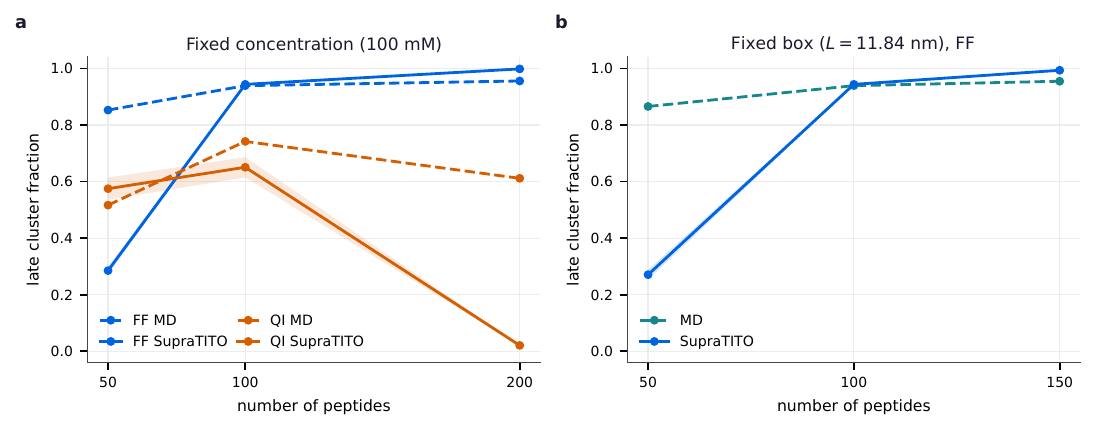}
  \caption{\textbf{Late structural response at new system sizes.}
  \textbf{a}, late
  largest-cluster fraction when peptide count and box size are changed to keep
  concentration near $100\,\mathrm{mM}$. \textbf{b}, FF peptide-count sweep
  at fixed box size. Dashed green lines show MD; solid blue lines and bands show
  the SupraTITO mean and one standard deviation across five independent
  rollouts.}
  \label{fig:system-size-late}
\end{figure*}

\begin{figure*}[!htbp]
  \centering
  \includegraphics[width=0.92\textwidth,height=0.72\textheight,keepaspectratio]{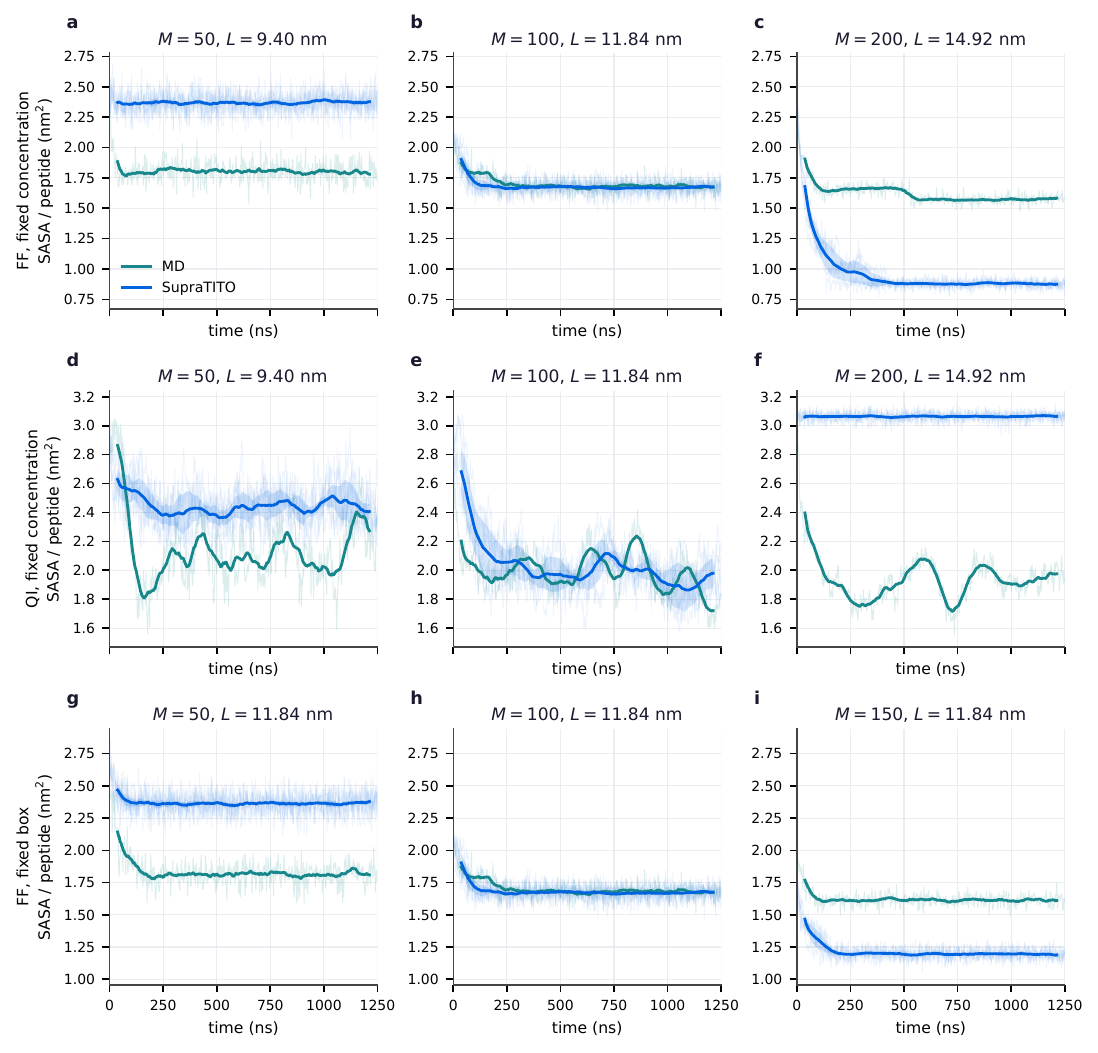}
  \caption{\textbf{Condition-specific SASA trajectories at new system
  sizes.} The top and middle rows vary peptide count and box size together at
  approximately $100\,\mathrm{mM}$ for FF and QI. The bottom row varies FF
  peptide count in a fixed $11.84\,\mathrm{nm}$ box. Every panel is one
  physical system. Line and uncertainty conventions follow
  \cref{fig:concentration-sasa-ag}.}
  \label{fig:system-size-sasa}
\end{figure*}

The fixed-concentration systems change peptide count and box size together,
whereas the fixed-box systems change peptide count at constant box length.
SupraTITO accepts the resulting variable number of retained sites, although it is
trained at only one system size. \Cref{fig:system-size-late} summarizes the
late structural response, and \cref{fig:system-size-sasa} shows the
corresponding time-resolved SASA.

Transfer across system size is less consistent than transfer across
concentration at fixed peptide number. At the smaller size, SupraTITO reproduces
the partial assembly of QI but leaves FF too dispersed. At the larger size, it
forms a nearly complete FF aggregate but does not recover the partially
assembled QI reference. The fixed-box FF sweep shows a similar sensitivity to
peptide number. These results separate architectural compatibility with a
variable number of sites from learned physical transfer: SupraTITO can be
evaluated at new sizes, but training at a single size does not determine all
finite-size effects on collective assembly.

\end{document}